\documentclass[11pt, twoside]{article}
\usepackage{graphicx}    % needed for including graphics e.g. EPS, PS
\usepackage{fancyhdr}
\usepackage{setspace}
\usepackage[all]{xy}
\usepackage{wrapfig}
\usepackage{amsfonts}
\fancyhfoffset[L]{ 0cm}
\title{Decoupling Gravity in F-Theory}
\author{Clay C\'{o}rdova}
\renewcommand{\thefootnote}{\fnsymbol{footnote}}
\begin{document}      
\thispagestyle{empty}   
% Start your text
\begin{center}
\Large{\textbf{Decoupling Gravity in F-Theory}} 
\end{center}
\vspace{-.125in}
\begin{center}
Clay C\'{o}rdova\footnote[1]{cordova@physics.harvard.edu} \\
\vspace{.165in}
\emph{Jefferson Physical Laboratory \\
Harvard University \\
Cambridge MA 02138}
\end{center}
\renewcommand{\thefootnote}{\arabic{footnote}}
\vspace{.15in}
\textbf{Abstract}: We study seven-brane \(SU(5)\) GUT models of string phenomenology which can be consistently analyzed in a purely local framework.  The requirement that gravity can decouple constrains the form of four-dimensional physics as well as the geometry of spacetime.  We rule out a large family of candidate UV completions of such models and derive a priori constraints on the local singularities of compact elliptic Calabi-Yau fourfolds.  These constraints are strong enough to obstruct a wide class of brane constructions from UV completion in string theory.    It is demonstrated that consistent local models always have exotic Yukawa coupling structures, and hidden sectors or interesting non-perturbative superpotentials which merit further investigation.
\begin{spacing}{.725}
\tableofcontents
\end{spacing}
\begin{center}
\section{Introduction}
\end{center}
\paragraph{}  
Recently, local brane models of string phenomenology have attracted significant attention.  A promising and broad class of such models is that of local F-theory GUTs \cite{BHVI} \cite{BHVII} \cite{Wijn}.  These scenarios provide a natural arena for supersymmetric grand unification, and lead to interesting phenomenologically viable constructions of dark matter, flavor, and neutrino physics \cite{M1} \cite{M2} \cite{M3} \cite{M4}.  In this setup, our four-dimensional world is realized as the non-compact directions of a stack of seven-branes which wrap a compact four-cycle \(S\) inside the ambient six-dimensional geometry \(X\) of the compactification.  Closed strings propagating in \(X\) give rise to gravitons, while open strings stuck to \(S\) produce the gauge bosons of the standard model.  Matter in the theory arises when a pair of seven-branes \(S\) and \(S^{\prime}\) intersect in \(X\) along some Riemann surface \(\Sigma\), a so-called matter curve.  There the quantization of open string modes starting on \(S\) and ending on \(S^{\prime}\) produces light matter localized on \(\Sigma\) which at low energies appears as quarks, leptons, neutrinos, and Higgses.  Finally, the superpotential of these theories is controlled by the triple intersection points of seven-branes in \(X\).  At these points three matter curves meet and an open string disk diagram with boundary at the point of intersection contributes a Yukawa coupling to the superpotential.
\begin{figure}[here]
\begin{center}
\includegraphics[totalheight=0.4\textheight]{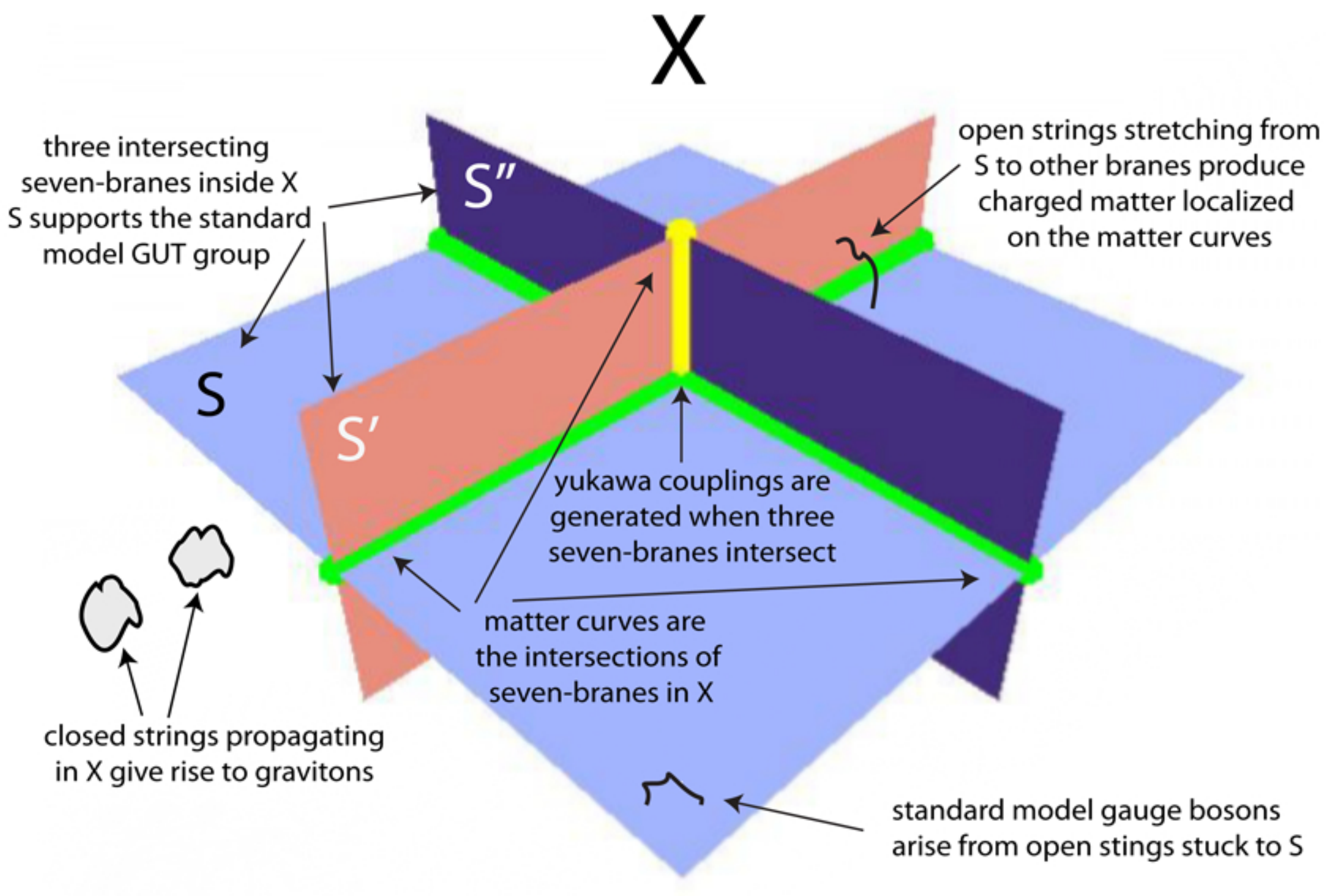}
\caption{The internal geometry of F-theory brane models and their various string sectors.}
\label{fig:intbrane}
\end{center}
\end{figure}
\paragraph{}
The basic feature of these models which makes them simpler than say, heterotic string phenomenology is that in the limit where the backreaction of seven-branes is ignored it is relatively straightforward to construct the standard model particle spectrum and interactions by simply prescribing geometrically the desired seven-brane intersections.  By contrast in compactifications of heterotic strings on Calabi-Yau threefolds it is often a difficult task to detect whether the resulting low energy physics in four-dimensions has anything to do with reality.  Of course the flipside to this discussion is that a heterotic compactification provides a UV complete theory including gravity, while in the case of local brane models, UV completion inside a compact threefold taking into account backreaction of seven-branes as well as coupling to the closed string sector is an involved geometry problem.
\paragraph{}
Thus, if we want to maintain the virtues of local brane models while avoiding their vices we are motivated to study situations where the limit of zero brane backreaction is likely to be a good starting approximation to the open string physics localized on the standard model seven-brane.    A simple way to achieve this is to demand that the brane models of interest have a decoupling limit where all interactions with gravity can be made parametrically small.    When such a limit exists one can reasonably hope that issues of gravitational physics and moduli stabilization can be deferred to a later stage of analysis without spoiling the particle physics features engineered in a local model.  Further there are suggestive hints from nature that a decoupling limit may be relevant for particle physics.  Indeed one can take the hugeness of the Planck scale, \(M_{P}\sim 10^{19}\) GeV, as quantitative proof that for most practical purposes of particle physics gravity does not play an essential role.  Going further, one might argue that the asymptotic freedom of the gauge coupling in a GUT model is evidence that the open string sector responsible for the standard model should be UV complete on its own, without necessarily coupling to gravity.  Whether or not one is convinced by these arguments suggesting the necessity of a gravitational decoupling limit in nature, models where gravity decouples certainly yield the simplest class of quasi-realistic string compactifications including branes.  What's more, these models are sufficiently rich that they can in principle accommodate even the most baroque features of the standard model.  Indeed, it is precisely in this limit that the phenomenological successes of Heckman, Vafa, and collaborators have been achieved.  
\paragraph{}
The technical power of the existence of a decoupling limit is that it implies a dramatic simplification of the brane geometry in Figure \ref{fig:intbrane}.  To understand this one must first appreciate that the four-dimensional GUT models of interest are described by three continuous parameters, the Planck scale \(M_{P}\), the GUT coupling \(\alpha\), and the GUT scale \(M_{GUT}\), each of which has a geometric interpretation in Figure \ref{fig:intbrane}.  The Planck scale is determined by dimensional reduction of the ten-dimensional Einstein-Hilbert action, so in ten-dimensional Planck units:
\begin{equation}
M_{P}^{2}\sim Vol(X) \label{Planck}
\end{equation}
Meanwhile the four-dimensional GUT coupling descends from the reduction of the Yang-Mills action on the seven-brane worldvolume:
\begin{equation}
\frac{1}{\alpha}\sim Vol(S) \label{coupling}
\end{equation}
Finally, the GUT scale is set by whatever mechanism Higgses the GUT group down to the usual \(SU(3)\times SU(2)\times U(1)\) of the standard model.  As part of our assumption of the existence of a decoupling limit, we will take as given that the physics responsible for the spontaneous breaking of the GUT group is adequately described by the gauge theory on the standard model seven-brane.  In this case the GUT scale is set by dimensional analysis:
\begin{equation}
M_{GUT}\sim Vol(S)^{-1/4} \label{GUT}
\end{equation}
If we want gravity to decouple in four-dimensions, we want to be able to take a limit where \(M_{P}\rightarrow \infty\), while the gauge theory parameters \(\alpha\) and \(M_{GUT}\) remain fixed.  Examining equations \((\ref{Planck})-(\ref{GUT})\) we see that geometrically this means that there should exist a limit where \(Vol(X)\rightarrow \infty\) while \(Vol(S)\) remains fixed.  This is a powerful geometric assumption, and investigating its consequences in detail forms the subject of this paper.
\paragraph{}
Although some work in the vein of global completions of F-theory GUTs with decoupling limits has already been carried out \cite{curio} \cite{Grimm} \cite{Donagi} \cite{Mar} \cite{Mar3}, a complete and consistent picture has not yet emerged.  We will focus primarily on the physical consequences of this limit which hold independent of a choice of compactification \(X\).  To the extent that we do discuss global properties of \(X\), we are interested mainly in learning what kinds of compactifications we are dealing with, and what information the existence of a decoupling limit implies about the properties of spacetime.  For the first part of this paper we give a brief review of the relevant geometry and study the simplest class of examples where the six-dimensional compactification manifold \(X\) is a Fano threefold.  As will hopefully be clear by the end of Section \ref{kahlersec}, such examples are completely ruled out.  We then move on in Section \ref{obstruc} to study what happens when the assumption of Fanoness is removed.  As we will see there, simple tadpole arguments together with the insight gained in Section \(\ref{kahlersec}\) are enough to obstruct any semi-realistic model with GUT group \(SU(5)\) and a simple decoupling limit from a UV completion in string theory.  In fact, the constraints on colliding seven-branes described in Section 4.1 are independent of the existence of a decoupling limit and represent a priori restrictions on the local singularities of any compact elliptically fibered Calabi-Yau fourfold.  When these constraints are combined with the decoupling limit hypothesis a surprising amount information about the general features of any local \(SU(5)\) F-theory GUT can be exposed.  With this in mind, in Section 4.1 we discuss the implications of our work for the recently engineered local F-theory GUTs in \cite{BHVI} \cite{BHVII} \cite{M1} \cite{M2} \cite{M3} \cite{M4}.  Building on previous work \cite{Donagi} \cite{BHVII}, we then finish in Section 4.2 by classifying all possible brane worldvolumes on which gravity can in principle decouple and begin to explore models with more exotic decoupling limits.  We conclude that the geometry of these exotic decoupling limits almost certainly plays a role in the local physics.
\paragraph{}
The intuitive idea of our arguments is to study the local gravitational backreaction of seven-branes on the geometry of the compactification.  From the point of view of general relativity a seven-brane is a rather subtle object.  Because they have only two transverse dimensions a seven-brane behaves like an isolated point mass in a three-dimensional spacetime.  For such a system, Einstein's equations imply that the effect of the point mass is so strong as to change the asymptotic shape of space into a cone with a deficit angle depending on the mass of the point.  So too it is with seven-branes.  In the supergravity approximation, an isolated, decompactified \(D7\) produces a conical deficit angle of \(\frac{\pi}{12}\) in the transverse dimensions, or said differently, a localized positive contribution to the Ricci curvature of \(X\).   Thus we see that there is conceptually a certain tension between, on the one hand, the desire to decouple gravity and deal only with the gauge theory supported on our seven-brane, and on the other hand, the fact that seven-branes produce a quite severe local gravitational backreaction.  This problem becomes particularly acute for a stack of seven-branes which support an \(SU(n)\) type gauge symmetry.  As we will see in Section \ref{kahlersec}, locally positive Ricci curvature acts as an obstruction to taking a decoupling limit.  In Section \ref{obstruc} we then demonstrate that lowering the local Ricci curvature sufficiently as to permit a simple decoupling limit essentially requires us to put an orientifold plane directly on top of our seven-brane and spoils the fact that the gauge group is \(SU(n)\).
\begin{center}
\section{Geometric Preliminaries} 
\label{geosec}
\end{center}
\paragraph{}
In this introductory section we review the relevant background material for the kind of geometrical problem we will be considering.  As usual supersymmetry singles out complex algebraic geometry as the relevant framework.  Thus, for example, in the following the words curve, surface, and threefold refer to complex manifolds of complex dimension one, two, and three respectively.  For additional background material the reader is referred to \cite{GH}.
\subsection{Geometry of Spacetime}
\label{geospace}
\paragraph{}
We are studying four-dimensional \(\mathcal{N}=1\), type IIB compactifications on a threefold \(X\) where the axio-dilaton varies throughout spacetime.  The fact that the string coupling is non-constant means that these compactifications are in general non-perturbative.  Such models are conveniently described in the language of F-theory \cite{ftheory}.  By viewing the axio-dilaton as the complex structure modulus of an elliptic curve, we can form a complex fourfold \(Y\) which is an elliptic fibration over \(X\).  By construction, the fibration \(Y\) admits a section which is simply the compact part of spacetime \(X\subset Y\).  The locus in \(X\) over which the elliptic fibration degenerates determines the positions of various stacks of seven-branes in \(X\).  The condition that the seven-brane tadpoles cancel while preserving four-dimensional \(\mathcal{N}=1\) supersymmetry implies that the fibered fourfold \(Y\) is Calabi-Yau.  A very useful mathematical construction for describing such compactifications is to present \(Y\) as a Weierstrass model.  To do this we first recall that an elliptic curve can be defined by a cubic equation in \(\mathbb{P}^{2}\).  Using local coordinates \((x,y)\) on a patch of \(\mathbb{P}^{2}\), we can always put this equation in the Weierstrass form:
\begin{equation}
y^{2}=x^{3}+fx+g \label{weier}   
\end{equation}
Where in the above \(f\) and \(g\) are numbers characterizing the shape of the torus.  The curve described by \((\ref{weier})\) is non-singular provided that it has a non-vanishing discriminant \(\Delta\) given by:
\begin{equation}
\Delta=4f^{3}+27g^{2} \label{deltadef}
\end{equation}
\paragraph{}
Now that we have a handle on a single elliptic curve, to form an elliptic fibration over \(X\) all we need to do is to let the coordinates \(x,y\), vary holomorphically over \(X\).  More precisely, we now take \(x,y\) to be local coordinates on suitable line bundles over \(X\).  Homogeneity of equation \((\ref{weier})\) tells us that if \(y\) is a coordinate on a line bundle \(3\mathcal{L}\), then \(x\) must be a coordinate on the line bundle \(2\mathcal{L}\).  To determine what \(\mathcal{L}\) actually is all we need to do is require that the fourfold \(Y\) is Calabi-Yau.  This means that there should be a never-zero holomorphic 4-form \(\omega_{Y}\) on \(Y\) and since \(Y\) is a fibration we can write \(\omega_{Y}\) as:
\begin{equation}
\omega_{Y}=\frac{dx}{y}\wedge \omega_{X} \label{wedge}
\end{equation}
Where in the above \(\omega_{X}\) is a holomorphic 3-form on \(X\) and hence transforms over \(X\) in the canonical line bundle \(K_{X}\).  Since \(\omega_{Y}\) transforms trivially, it follows that \(\mathcal{L}=-K_{X}\) and hence in equations \((\ref{weier})\) and \((\ref{deltadef})\), \(f\), \(g\), and \(\Delta\) are fixed holomorphic sections of \(-4K_{X}\), \(-6K_{X}\), and \(-12K_{X}\) respectively.  Conversely, one can view the Weierstrass presentation of \(Y\) as a recipe for constructing elliptic Calabi-Yaus.  Given a threefold \(X\) satisfying certain assumptions which we review below, one picks sections \(f\) and \(g\) of \(-4K_{X}\) and \(-6K_{X}\) and defines \(Y\) as the solution to equation \((\ref{weier})\).
\paragraph{}
As we have already mentioned above, one useful feature of the F-theory description is that the elliptic structure of \(Y\) encodes the places in spacetime where seven-branes are located.  To understand this all we need to recall is that an ordinary perturbative seven-brane is a magnetic source for the IIB axion.  Thus as one circles a seven-brane the complexified string coupling \(\tau\) undergoes a monodromy \(\tau \rightarrow \tau +1\).  In the F-theory description, \(\tau\) is the modulus of the elliptic fibers of \(Y\) and the fact that \(\tau\) has monodromy around seven-branes means that the associated elliptic fiber is singular exactly at the seven-brane.  This story can be generalized \cite{geosing}; in F-theory suitable monodromies of the elliptic modulus can be prescribed to engineer seven-brane gauge groups of \(A\), \(D\), and most notably \(E\) type.  For computational purposes, it is more useful to translate the monodromies of \(\tau\) into vanishing orders of the sections \(f\), \(g\), \(\Delta\) defining the Weierstrass model \((\ref{weier})\).  We already know that the locus where the elliptic fibration degenerates is exactly defined by the vanishing of the discriminant \(\Delta\).  The precise gauge group can then be deduced by further studying the vanishing orders of the defining sections as described in Table 1.
\begin{center}
\begin{table}[h]
\begin{center}
\begin{tabular}{|c||c|c|c|c|c|c|c|c|c|}
\hline
Group & None & \(A_{n-1}\) & \(A_{1}\) & \(A_{2}\) & \(D_{n+4}\) & \(D_{n+4}\) & \(E_{6}\) & \(E_{7}\) & \(E_{8}\) \\
\hline
\(\Delta\) & 0 & \(n\) & \(3\) & \( 4\) & \(n+6\) & \(n+6\) & \(8\) & \(9\) & \(10\) \\
\hline
\(f\) & \(\geq 0\) & \(0\) & \(1\) & \(\geq 2\) & \(2\) & \(\geq 2\) & \(\geq 3\) & \(3\) & \(\geq 4\) \\
\hline
\(g\) & \(\geq0\) & \(0\) & \( \geq 2\) & \(2\) & \(\geq 3\) & \(3\) & \(4\) & \(\geq 5\) & \(5\) \\
\hline
\end{tabular}
\end{center}
\caption{Seven-brane gauge groups indexed by the vanishing order of \(f\), \(g\), and \(\Delta\).  The notation \(\geq\) means that the corresponding section vanishes to order greater than or equal that indicated. }
\end{table}
\end{center}
\paragraph{}
One universal feature of these F-theory compactifications is that the corresponding elliptic Calabi-Yaus are singular whenever there is non-abelian gauge symmetry on some seven-brane somewhere in spacetime.  Geometrically the case of a single \((p,q)\) seven-brane, an \(A_{0}\) fiber in the notation of Table 1, is distinguished by the fact that the fourfold near such a brane is nonsingular even though the elliptic fiber degenerates.   The fact that non-abelian gauge symmetries are described by singular Calabi-Yaus points to another important fact: \emph{the non-abelian gauge symmetry on any given seven-brane is bounded above in rank}.  Technically the way this comes about is that in order to make sense of the physics on a singular Calabi-Yau, one is forced to resolve the singularity.  If the singularity is too large, i.e. if a seven-brane has a gauge group of too high rank, then the resolution will fail to be Calabi-Yau and hence the original singular Calabi-Yau will break supersymmetry \cite{MV}.  One can get an estimate of the actual rank of the biggest possible singularity as well a gain some intuitive feeling for its meaning by working locally in the supergravity limit.  Then each seven-brane contributes a conical deficit angle of \(\frac{\pi}{12}\) so certainly at most one could have twenty four \(D7\)-branes on top of each other before the local deficit angle exceeds \(2\pi\).  In fact the actually bound is smaller; on any given brane the discriminant can at most vanish to order ten so the largest possible simple factor of the total seven-brane gauge group is \(E_{8}\).  This relatively small upper bound on the size of a stack seven-branes in any theory with gravity should be contrasted with geometrically engineered brane theories where backreaction can be completely ignored i.e. when one considers F-theory on a local non-compact Calabi-Yau fourfold with no intention of embedding it in a compact geometry where the metric is a dynamical field.  Then there is no bound on the number of branes and it is easy to construct seven-brane gauge theories with arbitrary \(ADE\) gauge group. 
\paragraph{}
Although the F-theory fourfold provides a convenient picture for simultaneously encoding both seven-branes positions and the internal geometry of spacetime, for the purposes of investigating the geometric properties of of brane models with decoupling limits it is easier to work directly with the threefold \(X\).    The simple reason for this is that the decoupling limit can be phrased easily in terms of the K\"{a}hler geometry of the threefold \(X\), and while the fourfold \(Y\) is Calabi-Yau the elliptic directions have no dynamical metric degrees of freedom hence no useful K\"{a}hler structure.  Thus before turning to an analysis of decoupling limits in Section \ref{kappa} we will first gain some intuition about what sorts of threefolds \(X\) solve the equations of motion for F-theory.  To begin with, we should emphasize the basic fact that for threefolds which include seven-branes the geometries in question are no longer Ricci flat.  To understand what behavior to expect for the Ricci curvature of \(X\) it is again helpful to think in the perturbative IIB limit.  As we have already mentioned, \(D7\)-branes produce conical deficit angles which are positive contributions to Ricci curvature.  Meanwhile orientifold planes produce negative Ricci curvature of \(X\) localized at their worldvolume.  Since the orientifold planes only occupy a sum of surfaces in \(X\) we can then conclude that \(X\) should have non-negative Ricci curvature away from the surfaces occupied by the orientifold planes.  Now we dial up the string coupling to transition from IIB to F-theory.  This smooths the singular contributions to the Ricci curvature of \(X\) and in general since we lack a quantitative knowledge of the effective action for the gravitational degrees of freedom in F-theory, we cannot make any assertions about the pointwise behavior of the curvature.  Nevertheless we can control local curvature averages in the form of the first Chern class of \(X\).  A simple way to see this is to examine the seven-brane tadpole equations that follow from the Weierstrass model.  The discriminant \(\Delta\) is a sum of surfaces \(S_{i}\) defining the compact part of various seven-brane worldvolumes:
\begin{equation}
\Delta=\sum_{i}n_{i}S_{i} \label{disc1}
\end{equation}
Where the \(n_{i}\) in \((\ref{disc1})\) are determined from the gauge group on each seven-brane via Table 1.  We know from the Calabi-Yau condition that \(\Delta\) is a section of \(-12K_{X}\), and topologically \(-K_{X}\) represents the first Chern class of \(X\), \(c_{1}(X)\).  Now let \(C\) in \(X\) be any complex curve.  We have:
\begin{equation}
\int_{C}\mathrm{Ricci}(X)=c_{1}(X)\cdot C = \frac{1}{12}\sum_{i}n_{i} S_{i}\cdot C \label{sdotc}
\end{equation}
The right-had-side of \((\ref{sdotc})\) is an intersection of complex manifolds and hence is non-negative provided that \(C\) is not contained in any of the seven-brane worldvolumes \(S_{i}\).  Thus in analogy with the IIB case we find that \(c_{1}(X)\geq0\) away from the seven-brane worldvolumes.  One can also turn this condition around, curves in \(X\) for which \(c_{1}(X)\cdot C<0\) are always contained inside some seven-brane in \(X\).  In this way \(c_{1}(X)\) controls the number of seven-brane moduli with negativity of the first Chern class along some curves obstructing any hypothetical seven-brane deformation where these curves exit the branes.  
\paragraph{}
In fact, one can make a sharper statement about the relation between seven-brane moduli and negativity of \(c_{1}(X)\).  To begin with, suppose for purposes of illustration that we were interested in the six-dimensional gauge theories obtained by compactification of F-theory on Calabi-Yau threefolds.  Then \(X\) is a complex surface and seven-branes wrap complex curves inside \(X\).  The argument following equation \((\ref{sdotc})\) then implies that any curve \(C\) where \(c_{1}(X)\cdot C <0\) is always wrapped by some seven-brane.  But now the sections \(f\) and \(g\) entering the definition of the Weierstrass model are also topologically represented by positive powers of \(c_{1}(X)\).  Hence the line of reasoning which led us to conclude that a seven-brane wraps \(C\) also implies that both \(f\) and \(g\) vanish on \(C\).  Examining Table 1 we conclude that \(C\) actually supports a \emph{non-abelian} seven-brane.  Turning this argument around we see that if \(c_{1}(X)\) is negative on any curve, the model is obstructed from Higgsing the total seven-brane gauge group to an abelian group.  Now let's upgrade this argument to the more interesting case of F-theory on a Calabi-Yau fourfold elliptically fibered over a threefold \(X\).  The same logic now implies that any curve \(C\) on which \(c_{1}(X)\) is negative carries an enhanced singularity larger than a single \(U(1)\).  Thus either \(C\) is contained in a non-abelian brane or \(C\) is matter curve located at the intersection of two branes where the degeneration type of the elliptic fibration enhances.
\paragraph{}
Beyond the basic requirement of having \(c_{1}(X)\) positive away from some loci of branes, another technical requirement we will put on \(X\) is that the vanishing loci of \(f\), \(g\), \(\Delta\) should always fit into Table 1 so that we can make sense of the theory in terms of usual gauge theories instead of say some unknown exotic stringy physics.  The class of threefolds \(X\) which satisfy these requirements form a rather large and varied set of geometries, and a complete classification of such \(X\) is not known.  A significant complication is that we are explicitly interested in non-abelian gauge seven-branes and therefore singular Calabi-Yaus.  There is however one family of threefolds which can always be the base of an F-theory fourfold.  These are threefolds with \(c_{1}(X)\) positive everywhere, the so-called Fano threefolds.  Roughly speaking, positive first Chern class means that the bundles \(-nK_{X}\) for \(n>0\) which entered in the definition of the Weierstrass model have a large number of sections.  It follows that Fanos are compactifications with a large number of seven-brane moduli suffering from none of the interesting obstructions outlined in the previous paragraph.  In the sense that these moduli must eventually be stabilized in any complete model Fanos might be a bad starting point.  Nevertheless since Fano threefolds form a completely classified set of simple geometries we will use them as interesting examples in the following.  
\subsection{Introduction to Fano Threefolds}
\label{fano3int}
\paragraph{}
Since Fano threefolds may be unfamiliar to some readers, in this section we spend some time enumerating their various special properties.  For further information the reader is referred to \cite{fanoref}.  Fano threefolds are three-dimensional algebraic varieties which admit K\"{a}hler metrics with strictly positive Ricci curvature.  They are the three-dimensional analogue of the famous del Pezzo surfaces.  The positivity of the curvature is an extremely strong topological condition on manifold.  Geometrically the positive curvature forces geodesics to bend toward themselves in such a way that the manifold closes up quickly before anything too drastic has occurred.  Since we are working with algebraic varieties, it is more convenient to phrase the positivity of the curvature in terms of the first Chern class of \(X\).  Fano threefolds are then characterized by the fact that \(c_{1}(X)\) intersects positively with every curve in \(X\).  Yau's theorem \cite{Yau} guarantees that we can pass from a representative of \(c_{1}(X)\) to a positive curvature metric in any K\"{a}hler class so little information is lost by working at the level of cohomology.  Further constraints on the geometry of \(X\) can be obtained by applying Kodaira vanishing and Serre duality:
\begin{equation}h^{i}(X)=h^{3-i}(K_{X})=0 \hspace{.3in} i>0 \label{kodairav}
\end{equation}
Where in equation \((\ref{kodairav})\) the final result follows because \(K_{X}\), being represented by the negative class \(-c_{1}(X)\), possesses a metric with strictly negative curvature.  Thus we immediately learn that the Hodge diamond of a Fano threefold \(X\) takes the rather restricted form:
\begin{eqnarray}
\begin{tabular}{cccccc}
\multicolumn{6}{c}{\(h^{3,3}(X)\)}  \\
\multicolumn{6}{c}{\(h^{3,2}(X)\) \hspace{.15in} \(h^{2,3}(X)\)} \\
\multicolumn{6}{c}{\(h^{3,1}(X)\) \hspace{.15in} \(h^{2,2}(X)\) \hspace{.15in} \(h^{1,3}(X)\)}  \\
\multicolumn{6}{c}{\(h^{3,0}(X)\) \hspace{.15in} \(h^{2,1}(X)\) \hspace{.15in} \(h^{1,2}(X)\) \hspace{.15in} \(h^{0,3}(X)\)} \\
\multicolumn{6}{c}{\(h^{2,0}(X)\) \hspace{.15in} \(h^{1,1}(X)\) \hspace{.15in} \(h^{0,2}(X)\)} \\
\multicolumn{6}{c}{\(h^{1,0}(X)\) \hspace{.15in} \(h^{0,1}(X)\)} \\ 
\multicolumn{6}{c}{\(h^{0,0}(X)\)} 
\end{tabular}
& = & 
\begin{tabular}{cccccc}
\multicolumn{6}{c}{1}  \\
\multicolumn{6}{c}{0 \hspace{.175in} 0} \\
\multicolumn{6}{c}{0 \hspace{.15in} \(h^{1,1}(X)\) \hspace{.15in} 0} \\
\multicolumn{6}{c}{0 \hspace{.15in} \(h^{2,1}(X)\) \hspace{.15in} \(h^{2,1}(X)\) \hspace{.15in} 0} \\
\multicolumn{6}{c}{0 \hspace{.15in} \(h^{1,1}(X)\) \hspace{.15in} 0} \\
\multicolumn{6}{c}{0 \hspace{.175in} 0} \\ 
\multicolumn{6}{c}{1} 
\end{tabular} 
\label{hodged}
\end{eqnarray}
In particular we see that the number of equivalence classes of divisors in \(X\), \(h^{1,1}(X)\), together with the remaining Hodge number of \(X\), \(h^{2,1}(X)\), are topological invariants of \(X\).  In fact the Hodge diamond structure \((\ref{hodged})\) must hold more generally for any \(X\) which forms the base of an elliptically fibered Calabi-Yau fourfold \(Y\) of full \(SU(4)\) holonomy.  Indeed if \(X\) had a nontrivial holomorphic form then this form would pullback to the fourfold contradicting the fact that \(h^{i,0}(Y)\) vanishes for \(i<4\).
\paragraph{}
Perhaps the most surprising feature of Fano threefolds is that unlike Calabi-Yau threefolds there are very few of them.  One can at least partially understand this fact by thinking about the situation for complex surfaces.  Suppose \(D\) is a del Pezzo surface, that is a two-dimensional complex variety which admits a K\"{a}hler metric with strictly positive Ricci curvature.  Then applying Kodaira vanishing as in \((\ref{kodairav})\) shows that the topological Euler characteristic of \(D\), \(\chi_{\hspace{-.05in}\phantom{a}_{Top}}(D)\), and the holomorphic Euler characteristic of \(D\), \(\chi_{\hspace{-.05in}\phantom{a}_{Hol}}(D)\), are given by:
\begin{equation}
\chi_{\hspace{-.05in}\phantom{a}_{Top}}(D)= 2+h^{1,1}(D) \hspace{.5in} \chi_{\hspace{-.05in}\phantom{a}_{Hol}}(D)=1 \label{eulerc}
\end{equation}
Now apply the index theorem: 
\begin{equation}
\chi_{\hspace{-.05in}\phantom{a}_{Hol}}(D)=\frac{1}{12}\left(c_{1}(D)^{2}+\chi_{\hspace{-.05in}\phantom{a}_{Top}}(D)\right) \label{indexs}
\end{equation}
Combined with equation \((\ref{eulerc})\) this yields:
\begin{equation}
c_{1}(D)^{2}+h^{1,1}(D)=10 \label{hodgesum}
\end{equation}
And since both terms on the left-hand-side of \((\ref{hodgesum})\) are positive, we learn that the cohomological possibilities for \(D\) are very limited.  Of course as is well known there are exactly ten del Pezzo surfaces given by \(\mathbb{P}^{1}\times \mathbb{P}^{1}\), \(\mathbb{P}^{2}\), or \(\mathbb{P}^{2}\) blown up at no more that eight points in general position.  The classification of smooth Fano threefolds is significantly more complicated.  Nevertheless a complete classification was obtained by Mori and Mukai in the 80's  \cite{fano1} \cite{fano2}, and as with del Pezzo surfaces the allowed values for \(h^{1,1}(X)\) are very limited.  There are exactly 105 possibilities for \(X\) with the number of each allowed \(h^{1,1}(X)\) listed below:
\begin{center}
\begin{tabular}{|c|c|c|c|c|c|c|c|c|c|c|c|}
\hline
\(h^{1,1}(X)\) & 1 & 2 & 3 & 4 & 5 & 6 & 7 & 8 & 9 & 10 & \(\geq 11\)  \\
\hline
\( \#\) & 17 & 36 & 31 & 13 & 3 & 1 & 1 & 1 & 1& 1 & 0\\
\hline
\end{tabular}
\end{center}
Clearly in order to admit any kind of decoupling limit \(X\) must have at least two independent scales \(Vol(X)\) and \(Vol(S)\) so \(h^{1,1}(X)> 1\).  Since the possibilities for \(X\) are so few, this constraint is in fact fairly non-trivial, ruling out a reasonable fraction of candidate threefolds.  
\paragraph{}
Although the complete classification of Fano threefolds is rather involved, the key ideas are simple to explain and relevant to the geometry in the rest of the paper.  As a warm up, let us first recall the classification of del Pezzo surfaces \(D\).  This is achieved by studying the K\"{a}hler cone.  Beyond its dimension, the interesting feature of any K\"{a}hler cone is its boundary, which describes possible degenerations of the del Pezzo where the metric fails to be positive.  The basic structure theorem for del Pezzos is that the faces on the boundary of K\"{a}hler cone, where a single cohomology class of curves shrinks to zero volume, always correspond to shrinking a \(\mathbb{P}^{1}\) inside the del Pezzo.  It is now a short step to see that, aside from the trivial case of \(D=\mathbb{P}^{1}\times \mathbb{P}^{1}\), these faces describe the elementary algebraic operation of a blowdown, and further that a blowdown of a del Pezzo remains del Pezzo.  Turning this idea around we find that to classify del Pezzo surfaces it suffices to find del Pezzos with a single K\"{a}hler class, and then study their blowups.  Since the only del Pezzo surface with \(h^{1,1}(D)=1\) is \(\mathbb{P}^{2}\) one then obtains all del Pezzos by blowing up \(\mathbb{P}^{2}\).
\paragraph{}
To upgrade this approach to the classification of Fano threefolds one needs first to understand the boundary of the K\"{a}hler cone of such threefolds.  This was achieved by Mori \cite{ConeThm} who classified all degenerations of Fano threefolds where the class of a single surface \(S\) shrinks to zero volume.  The del Pezzo surface \(\mathbb{P}^{1}\times \mathbb{P}^{1}\) is now replaced by threefolds which are either fibrations of del Pezzo surfaces over \(\mathbb{P}^{1}\), or \(\mathbb{P}^{1}\) fibrations over del Pezzo surfaces.\footnote{These are algebraic fibrations, so degenerations of the fibers generically occur.}  Meanwhile the operation of blowing down a \(\mathbb{P}^{1}\) inside a del Pezzo is replaced by contractions of surfaces of four possible types:
\begin{itemize}
\item \(S\) is \(\mathbb{P}^{2}\) which shrinks to a point.  This case includes the familiar construction of blowing up a threefold at a point.
\item \(S\) is \(\mathbb{P}^{1}\times \mathbb{P}^{1}\) which shrinks to a point.  The two \(\mathbb{P}^{1}\) directions on the surface are always cohomologically equal in \(X\).
\item \(S\) is a singular cone which shrinks to a point.  This cone is defined in projective coordinates in \( \mathbb{P}^{3}\) by the equation \(x^{2}+y^{2}+z^{2}=0\).
\item \(S\) is a \(\mathbb{P}^{1}\) fibration over a smooth curve \(C\) which shrinks to the curve \(C\) by collapsing the \(\mathbb{P}^{1}\) fibers.
\end{itemize}
To complete the classification of Fano threefolds following the example of del Pezzo surfaces, one next classifies simple cases where \(h^{1,1}(X)=1\) or \(2\) and hence the K\"{a}hler cone has little interesting boundary structure.  Finally, one then studies how to produce new Fanos from these simpler ones by blowing them up along points and curves creating the surfaces \(S\) appearing on the above list.  
\paragraph{}
The principle fact that the reader should take away from this discussion is that on Fano threefolds, classifying the basic allowed K\"{a}hler degenerations where a single surface \(S\) shrinks is a completely understood problem.  As we will discuss in the next section the gravitational decoupling limit that we want to take is essentially a K\"{a}hler degeneration so understanding this list together with a bit about Mori's method will take us a long way towards ruling out Fano threefolds as candidate UV completions of local F-theory GUTs.
\begin{center}
\section{Decoupling Limits and Fano Threefolds}
\label{kahlersec}
\end{center}
\paragraph{}
Now we turn to a more detailed study of the decoupling limit.  It is useful to divorce two conceptually distinct issues.  The first is the study of complex surfaces \(S \subset X\) on which it is in principle possible to wrap a seven-brane of any type and take a decoupling limit.  The second is an analysis of an actual gauge theory on such a surface \(S\).  As we will see in this section only the former is actually relevant for ruling out Fano threefolds as candidate compactifications.  Thus throughout the remainder of this section the reader will find almost no mention of any properties of gauge theories, only a geometric analysis of decoupling limits.
\subsection{Decoupling Limit Geometry}
\label{kappa}
\paragraph{}To setup the problem we will first introduce a convenient geometric picture for thinking about the decoupling limit.  As we have already discussed in the introduction, a necessary condition for decoupling gravity on a seven-brane wrapped on \(S\) is that we can take a limit where the Planck mass becomes large while the gauge theory parameters stay fixed.  Using the estimates given in \((\ref{Planck})-(\ref{GUT})\) this means that we can take a limit where \(Vol(X)\) becomes parametrically large while \(Vol(S)\) remains fixed.  It is mathematically convenient to rewrite this requirement as follows.  First we use our estimates to deduce that:
\begin{equation}
\frac{M_{GUT}}{M_{P}}\sim\alpha\left( \frac{Vol(S)^{3/4}}{Vol(X)^{1/2}}\right)
\end{equation}
Hence for fixed gauge coupling constant \(\alpha\), the existence of a decoupling limit means that the ratio \(\frac{Vol(S)^{3/4}}{Vol(X)^{1/2}}\) can be made parametrically small.
Slightly more formally, the existence of a decoupling limit implies that we have a one-parameter family of K\"{a}hler classes \(\omega(t)\) with the property that:
\begin{equation}
\lim_{t\rightarrow \infty} \ \  \frac{(\int_{S}\omega(t)^{2})^{3/4}}{(\int_{X}\omega(t)^{3})^{1/2}}= 0 \label{tpar}
\end{equation} 
Written in this form it obvious that the decoupling condition is insensitive to the overall normalization of the K\"{a}hler class.  Given any path \(\omega(t)\) in the K\"{a}hler cone satisfying \((\ref{tpar})\), we can obtain another such path by multiplying \(\omega(t)\) by any positive real function \(f(t)\).  Physically the decoupling limit is a process where the volume of \(X\) becomes very large with the size of the surface \(S\) fixed. However mathematically this is inconvenient because in the limit we are forced to deal with a non-compact threefold.  We will thus find it more useful to analyze the geometry of the decoupling limit by renormalizing \(\omega(t)\) such that a finite non-zero limiting class \(\omega\) exists.\footnote{Strictly speaking, to ensure the existence of a limit one may have to pass from the one-parameter family \(\omega(t)\) to a sequence of classes \(\omega_{i}\) and then finally to a subsequence with a limit.  This small subtlety together with additional details about this construction are explained more fully in Appendix A.}  The decoupling condition \((\ref{tpar})\) then implies that \(S\) has zero volume as measured by the limit class \(\omega\):  
\begin{equation}
\omega^{2}\cdot S = 0
\end{equation} 
\paragraph{}
Geometrically the reason that a finite limit class \(\omega\) is useful is that \(\omega\) lies on the boundary of the K\"{a}hler cone of \(X\) so this rescaled version of the decoupling limit is now simply a K\"{a}hler degeneration of \(X\), and can be analyzed using familiar techniques of algebraic geometry.  In particular we can make sense of the limit of \(X\) itself as some compact complex manifold by simply defining \(\omega\) to be a K\"{a}hler class on the limit.  Notice that it is key for this construction that we are studying a \emph{relative} K\"{a}hler degeneration where the volume of \(X\) becomes large while the volume of the surface \(S\) is fixed.  This is not the most general kind of limit in K\"{a}hler moduli space.  For example in the context of mirror symmetry of type II strings on Calabi-Yau threefolds one might be interested in the mirror of a large complex structure limit which would correspond to taking the K\"{a}hler class towards infinity with no fixed reference volumes.  In this case nothing can be gained by renormalizing the K\"{a}hler class.
\paragraph{}
Qualitatively speaking, there are now two possibilities depending on whether \(\omega^{3}\) is or is not equal to zero.  When \(\omega^{3}\) vanishes, the limit of \(X\) has zero three-dimensional volume and so \(X\) itself has also collapsed to a surface or a curve.  Geometrically this means that at least one dimension of \(S\) spans a dimension of \(X\) so that when \(S\) shrinks \(X\) is also forced to shrink.  The simplest examples of this type are when asymptotically, as we approach the decoupling limit, \(X\) looks like a fibration of \(S\) over a curve \(C\).    More generally \(X\) may not be a fibration, but it still admits a holomorphic map to surface or a curve with the brane worldvolume \(S\) collapsed by this map.  One can most likely study these models by investigating F-theory on elliptically fibered Calabi-Yau surfaces or curves, and then fibering these over curves or surfaces respectively.  To understand this, let's stick for the moment with the case where \(X\) degenerates to a curve.  A typical surface, \(F\), collapsed by this degeneration will then have trivial normal bundle in \(X\).  The adjunction formula then implies that:
\begin{equation}
K_{F}=K_{X}|_{F}+N_{F/X}=K_{X}|_{F}
\end{equation}
So the canonical bundle of \(F\) is simply the canonical bundle of \(X\) restricted to \(F\).  It follows that if we restrict the Weierstrass model of \(Y\) to \(F\), we obtain an elliptically fibered Calabi-Yau threefold.  The full Calabi-Yau fourfold \(Y\) is then obtained by fibering these threefolds over the curve \(C\) which is in fact a \(\mathbb{P}^{1}\).  A similar story holds for the case when the decoupling limit is such that \(X\) degenerates to a surface.  The Weierstrass model restricted to a typical fiber then gives an elliptic \(K3\) with the full Calabi-Yau fourfold constructed by fibering these \(K3\)'s over a complex surface \(D\) with \(h^{1,0}(D)=h^{2,0}(D)=0\).  Strictly speaking, the most general class of models with \(\omega^{3}=0\) are not simply fibrations; the gluing of the lower dimensional Calabi-Yau's together may involve interesting subtleties at special fibers of the decoupling map.  Nevertheless the main point of this analysis should be clear: models with \(\omega^{3}=0\) are geometrically degenerate in that they are glued together out of lower dimensional Calabi-Yau's.  Thus although there is nothing physically wrong with these constructions, for the remainder of the paper we will focus on the more interesting decoupling limits where \(\omega^{3}\neq0\).  In this case the limit of \(X\) also looks three-dimensional and one learns the least amount of global information about the Calabi-Yau fourfold.  Geometrically these are certainly the most robust examples to study and they include all local models discussed in the literature to date \cite{BHVII} \cite{Grimm} \cite{Grimm2} \cite{Donagi} \cite{Mar}.
\paragraph{}
The decoupling limits of interest are thus K\"{a}hler degenerations of the threefold \(X\) where the compact part of our \(SU(5)\) brane worldvolume \(S\) collapses while leaving the bulk size of \(X\) at finite volume.\footnote{As discussed in Appendix A this means mathematically that the decoupling limit is described by a \emph{birational} transformation of the threefold.}  In this case we can draw the conclusion that \(S\) must be a rigid cycle which admits no holomorphic normal deformations in \(X\).  The reason for this is simply that on the K\"{a}hler manifold \(X\) the volume of the holomorphic cycle \(S\) depends only on the \emph{cohomology class} of \(S\) in \(X\).  In particular any normal deformation of \(S\), being cohomologous to \(S\) itself, has the same volume as \(S\).  Thus if \(S\) were non-rigid, collapsing \(S\) would require collapsing the three-dimensional region spanned by \(S\) and its normal deformations and hence would collapse the threefold \(X\) itself.  
\begin{wrapfigure}{r}{.5\textwidth}
\fbox{
\begin{minipage}{3.05in}
\begin{center}
\includegraphics[totalheight=0.4\textwidth]{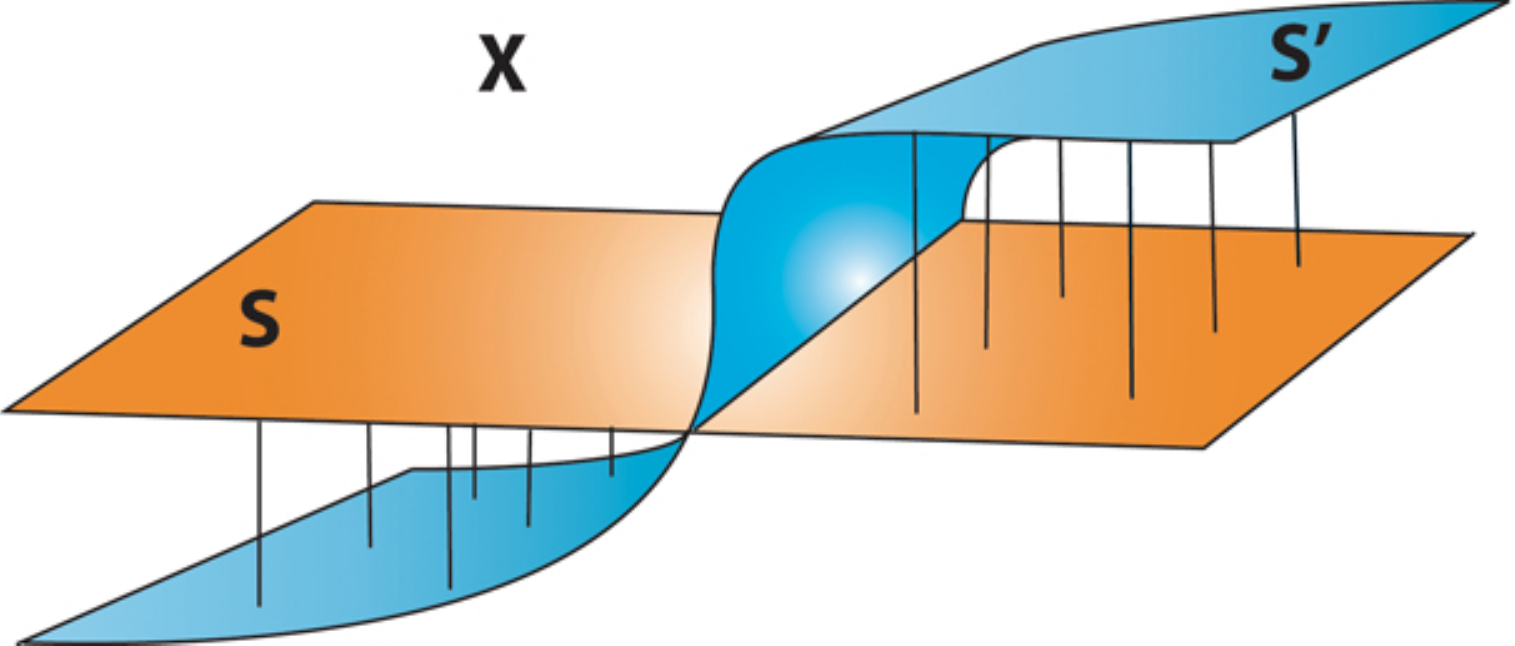}
\end{center}
\caption{A non-rigid cycle \(S\) whose collapse would lead to the collapse of the three-dimensional striated region.}
\end{minipage}
}
\end{wrapfigure}
Physically the spectrum of holomorphic normal deformations of \(S\) in \(X\) is realized in the effective four-dimensional GUT theory supported on the seven-brane as adjoint scalar fields.  The fact that the cycle \(S\) must be rigid then tells us the interesting fact that there are no such four-dimensional adjoints.  From the point of view of GUT models this is rather intriguing; the traditional mechanism for Higgsing the GUT group to the gauge group of the standard model relies precisely on giving an adjoint Higgs a suitable expectation value.  What we are then learning is that in F-theory GUTs with decoupling limits this mechanism is unavailable and hence we must utilize a mechanism such as brane flux or Wilson lines intrinsic to the higher-dimensional nature of the GUT theory \cite{BHVII}.
\paragraph{}
Actually there are two important subtleties in the discussion of the spectrum of the theory on the seven-brane.  On a flat seven-brane, the bosonic fields in the eight-dimensional \(\mathcal{N}=1\) gauge theory are an eight-dimensional gauge field, \(A_{\mu}\), as well as an adjoint complex scalar, \(\phi\), whose expectation values parameterize normal motions of brane.  To preserve supersymmetry when we compactify this theory on the K\"{a}hler surface \(S\) the theory must then be topologically twisted along the compact dimensions.  Curiously, once the surface \(S\) is specified, there is a unique supersymmetric twist available and after the twist \(\phi\) now transforms as a holomorphic two-form on \(S\) while \(A_{\mu}\) retains its spin \cite{BHVI}.  Now we reduce to the effective four-dimensional action and we see that there are two distinct sources of adjoint chiral superfields in the theory.  The first from the reduction of \(\phi\) yields \(h^{2,0}(S)\) multiplets, while the second from the reduction of \(A_{\mu}\) yields \(h^{1,0}(S)\) such multiplets.  What distinguishes these two classes of adjoints in the effective four-dimesnional theory is their \emph{couplings} i.e. the way that they enter the superpotential.  Holomorphy of the superpotential then protects this distinction at least till the supersymmetry breaking scale where we expect all allowed operators to be generated by quantum corrections. We will discuss this further in Section 4.  However for now we can state a basic fact that in the absence of adjoints associated to the \(\phi\) field, the adjoints descending from the vector do not have a sufficient superpotential to Higgs the SU(5) GUT group \cite{BHVII}.
\paragraph{}
The second subtlety in this discussion is now manifest from the previous paragraph: the twist of the seven-brane gauge theory does not appear sensitive to the normal bundle of \(S\) in \(X\).  Thus in fact the claim that the holomorphic normal deformations of \(S\) are realized as four-dimensional chiral adjoints is strictly speaking not true.  As we have just discussed, the spectrum of four-dimesnional adjoints capable of Higgsing the GUT group is controlled by the number of holomorphic sections of the canonical bundle of \(S\).  This then presents a paradox: what four-dimensional modes do holomorphic normal deformations of \(S\) in \(X\) describe?  As we will see in Section 4, the answer to this puzzle is that the seven-brane tadpole equations connect the bundle \(K_{S}\) and \(N_{S/X}\) in such a way that the holomorphic sections of \(K_{S}\) are always a subset of the holomorphic sections of \(N_{S/X}\).  Roughly speaking what is happening is that some of the normal deformations of \(S\) in \(X\) are massive in the seven-brane gauge theory because one cannot consistently extend these deformations to the other seven-branes in X while maintaining supersymmetry.  The fact that the adjoints transform as sections of \(K_{S}\) and not \(N_{S/X}\) is simply encoding this fact.  In any case, the fundamental conclusion that a decoupling limit implies no four-dimensional adjoint Higgsing of the GUT group remains valid.
\paragraph{}
Now that we have discussed the basics of the decoupling limits in question, it is useful to make a more refined classification of the local picture of \(X\) near \(S\) when \(S\) collapses.  There are three qualitatively distinct possibilities:
\[
\begin{tabular}{l l l}
1. \emph{Elementary Contraction to a Point}: & & \(S\) shrinks to a point and no other surface shrinks.\\
& & \\
2.  \emph{Elementary Contraction to a Curve}: & &  \(S\) shrinks to a curve and no other surface shrinks. \\
& & \\
3. \emph{Non-Elementary Contraction}:  & &  S shrinks either to a point or a curve and another  \\
& & surface that meets \(S\) also shrinks.
\end{tabular}
\]
Both physically and mathematically the decoupling limits associated to elementary contractions are the cleanest.  In the case of a non-elementary contraction one expects that wrapped branes on the additional surfaces shrinking in the decoupling limit contribute to the effective four-dimensional spectrum and interactions for the GUT theory supported on \(S\).  Mathematically, elementary contractions are simplest because a piece of technology, Grauert's criterion \cite{Grauert}, gives necessary and sufficient conditions for them to occur.  The statement is that \(S\) can undergo an elementary contraction to a point if and only if \(c_{1}(N_{S/X})\) is negative along every curve in \(S\).  Similarly \(S\) can undergo an elementary contraction to a curve if \(c_{1}(N_{S/X})\) is negative along each fiber of the contraction of \(S\) to the limit curve.  
\paragraph{}
Given any particular threefold \(X\), it is straightforward to apply Grauert's criterion to determine which surfaces can undergo elementary contractions.  For example, consider the well studied case of type II strings on a Calabi-Yau threefold \(X\).   If \(S\subset X\) is a surface which can undergo an elementary contraction to a point then via the adjunction formula:
\begin{equation}
0=c_{1}(X)|_{S}= c_{1}(S)+c_{1}(N_{S/X})\label{CYDP}
\end{equation}
By Grauert, \(c_{1}(N_{S/X})\) is negative so equation \((\ref{CYDP})\) implies that \(c_{1}(S)\) is positive and hence \(S\) is del Pezzo.  Notice in this computation how significantly the canonical bundle of the threefold \(X\) entered.  An F-theory compactification is not a Calabi-Yau compactification and there are an infinite number of possible surfaces which can shrink inside \(X\) and form candidate brane worldvolumes.  Indeed if we put no restrictions on \(c_{1}(X)\) then there are no restrictions on the kinds of surfaces which can shrink.\footnote{To illustrate this point let \(S\) be any K\"{a}hler surface, and pick \(H\) a positively curved line bundle on \(S\).  Now form the K\"{a}hler threefold \(\mathbb{P}(\mathcal{O}_{S}(-H) \oplus \mathcal{O}_{S})=X\).  Then \(X\) is fibered over \(S\) and \(S\) sits inside \(X\) as a section.  Since we chose \(H\) to be positively curved, \(S\) has negative normal bundle in \(X\) and by Grauert \(S\) can be shrunk to a point inside \(X\).}  Of course as discussed in Section 2 there are restrictions on \(c_{1}(X)\) and in Section 4 we will analyze the resulting restrictions on placed on brane worldvolumes \(S\).  However for now we simply wish to make the point that in an F-theory compactification there is nothing a priori special about del Pezzo surfaces.
\subsection{A No-Go Result Against Fanos}
\label{nogosec}
\paragraph{}
In the previous section we outlined the basic geometry relevant to studying decoupling limits on seven-branes in any ambient threefold geometry \(X\).  Now we will restrict to Fano threefolds, so \(c_{1}(X)\) is positive definite.  The discussion at the end of Section 2 about Mori's classification of K\"{a}hler degenerations of Fano threefolds states that the only surfaces \(S\) in a Fano \(X\) which can undergo elementary contractions are \(\mathbb{P}^{2}\), quadric cones, and \(\mathbb{P}^{1}\) fibrations over curves.  The quadric cone is a singular complex surface and a formalism for investigating the four-dimensional gauge theories obtained by compactifying seven-branes on such surfaces has not yet been developed.  Thus in the following we will restrict ourselves to the smooth possibilities for \(S\).  
\paragraph{}
Now we will impose a single constraint which we believe is necessary for the phenomenological success of any seven-brane model.  We will demand that \(S\) has a sufficient number of candidate matter curves to construct a model which is remotely MSSM-like.  As we have reviewed in the introduction, Yukawa couplings are generated at points in \(S\) where matter curves intersect.  Given any pair of matter curves in \(S\) they will generically intersect for dimensional reasons and hence the existence of Yukawa couplings is enforced topologically by intersection theory.  To  geometrically engineer the structure of the MSSM interactions while at the same time avoiding such disasters as proton decay then requires a sufficient number of linearly independent cohomology classes of curves.  At the bare minimum we expect that \(h^{1,1}(S)\geq3\), leaving no possibilities for \(S\) in the case of elementary contractions inside Fanos.
\paragraph{}
The line of reasoning in the above paragraph still leaves open the possibility that \(S\) might undergo a non-elementary contraction inside a Fano \(X\).  Indeed the boundary of the K\"{a}hler cone of \(X\) contains not only the faces which describe degenerations classified by Mori, but also edges where faces intersect and multiple surfaces collapse yielding a non-elementary contraction.  At such an edge not only do the individual surfaces corresponding to each face collapse, but also every surface cohomologically equivalent to an arbitrary sum of these surfaces also shrinks.  To completely rule out Fano threefolds as candidate compactifications we thus need to generalize slightly Mori's analysis to account for this possibility.  
\paragraph{}
As a prerequisite we need to be more specific about what exactly the existence of a decoupling limit implies.  To this end, in Appendix B we prove a well known mathematical result called \emph{negativity of contraction}.  If \(S\) can shrink then necessarily \(S\) contains a curve \(C\) with \(c_{1}(N_{S/X})\cdot C <0 \).  And furthermore, the curve \(C\) deforms in \(S\), \(C\cdot C \geq0\).  As explained in Appendix B this implies that no positive power of the normal bundle of \(S\) admits any holomorphic sections so \(S\) is certainly rigid.  Now to analyze possible non-elementary contractions inside a Fano \(X\), we use the adjunction formula to relate the canonical bundle of \(S\) to the normal bundle of \(S\):
\begin{equation}
K_{S}=K_{X}|_{S}\otimes N_{S/X} \label{adjf}
\end{equation}
The canonical bundle of \(X\) is negative-definite since \(X\) is Fano, and by negativity of contraction we can find a deformable curve \(C\) on which \(N_{S/X}\) is negative.  Thus by \((\ref{adjf})\) we learn that the canonical bundle \(K_{S}\) is itself negative along the deformable curve \(C\).  The same logic from Appendix B that implies that no positive power of \(N_{S/X}\) admits sections then implies that no positive power of \(K_{S}\) admits sections.  Complex surfaces satisfying this property have been completely classified \cite{GH} and are known as ruled surfaces.  They are all of the following form: a \(\mathbb{P}^{1}\) bundle over an arbitrary curve of any genus, blown up at points an arbitrary number of times.  
\paragraph{}
To the result of the previous paragraph we now add our phenomenological restriction: \(h^{1,1}(S)\geq3\).  This means that there is at least one blowup of the ruled surface \(S\).  What we will now demonstrate is that as long as \(X\) is Fano, \emph{no such surface can ever contract}.  The complete proof of this fact is somewhat involved and relegated to Appendix C, however we can present simple cases to illustrate why this is so.  To begin with consider for example the case where \(S\) is ruled over a surface of genus zero.  In this case \(S\) is a blowup of a Hirzebruch surface \(\mathbb{F}_{n}\) at a positive number of points.   One interesting feature that all such surfaces have in common is that they have a Mori cone of curves which is spanned by rational curves (\(\mathbb{P}^{1}\)s) with strictly negative normal bundle.  That is if \(S\) is any blowup of a Hirzebruch surface, and \(C \subset S\) any curve then in terms of linear equivalence:
\begin{equation}
C=\sum_{i}a_{i}\Gamma_{i} \label{MORICONEC}
\end{equation}
Where in \((\ref{MORICONEC})\), each of the coefficients \(a_{i}\) is non-negative, and the curves \(\Gamma_{i}\) are rational curves with \(\Gamma_{i}\cdot \Gamma_{i}<0\).  This property of such \(S\) might sound rather esoteric, but as we will now show together with Fano condition this obstructs any such \(S\) from admitting even a non-elementary contraction.  To see this we need only apply the genus formula together with adjunction to each of the rational curves \(\Gamma_{i}\):
\begin{equation}
0=g(\Gamma_{i})=\frac{1}{2}(\Gamma_{i}-c_{1}(S))\cdot \Gamma_{i}+1=\frac{1}{2}(c_{1}(N_{S/X})-c_{1}(X))\cdot \Gamma_{i} +\frac{\Gamma_{i}\cdot \Gamma_{i}}{2}+1 \label{genusf}
\end{equation}
Since we are assuming that \(X\) is Fano and we know that \(\Gamma_{i}\cdot \Gamma_{i}<0\) equation \((\ref{genusf})\) implies that \(N_{S/X}\cdot \Gamma_{i}\geq 0\).  Together with the fact \((\ref{MORICONEC})\) that such \(\Gamma_{i}\) span the Mori cone, this argument shows that \(N_{S/X}\cdot C\geq0\) for any curve \(C\subset S\).  Comparing this with our discussion of negativity of contractions we conclude that a blowup of \(\mathbb{F}_{n}\) can never contract inside Fano threefolds. 
\paragraph{}
Another way to gain intuition for the geometric content of this no-go theorem is to attempt to build a counterexample.  As a strategy for trying to build a counterexample we start first with the degenerated limit of our threefold and then modify it by a sequence of blowups.  The exceptional divisors of the blowups then yield surfaces inside the modified threefold which admit decoupling limits.  The basic Chern class identities for blowups tell us that if \(\widetilde{Z}\) is a blowup of the threefold \(Z\) then the Chern classes are related by:
\begin{equation}
c_{1}(\widetilde{Z})=c_{1}(Z)-D \label{chernblow}
\end{equation}
Where in the above \(D\) is an effective divisor.  Thus, at least for curves not contained in the divisor \(D\), blowing up reduces the first Chern class.   Now we know from Mori's classification of elementary contractions on Fanos that the only smooth surfaces which collapse in a single blowdown are \(\mathbb{P}^{2}\) and minimal ruled surfaces so any hypothetical counterexample involves at least two blowups.  Since we want the end result of the blowups to be Fano, equation \((\ref{chernblow})\) suggests that we take as the degenerated limit of \(X\) a threefold with a large first Chern class.  Thus we will start with \(\mathbb{P}^{3}\) and attempt to create a Fano \(X\) with a decoupling limt surface \(S\) by blowing up twice.  
\paragraph{}
One such attempt is illustrated below in Figure 3.  We consider a singular curve \(C\) in \(\mathbb{P}^{3}\) with an ordinary double point at \(p \in \mathbb{P}^{3}\).  We blowup \(\mathbb{P}^{3}\) at \(p\) to create a \(\mathbb{P}^{2}\) denoted by \(E\) in the illustration.  The fact that the two branches of \(C\) meet at \(p\) with distinct tangents means that in the blowup, the strict transform of \(C\) meets \(E\) transversally at a pair of points.  Now we blowup again at the strict transform of \(C\) to obtain the final threefold \(X\).  In \(X\), \(E\) has been modified by a blowup at the two points of intersection with the strict transform of \(C\) creating a non-minimal del Pezzo, \(dP_{2}\).  Inside \(X\) this \(dP_{2}\) is contractible via a non-elementary contraction where \(X\) undergoes a sequence of blowdowns back to \(\mathbb{P}^{3}\).  Unfortunately however, \(X\) is not Fano.  Inside \(E\) there is a distinguished line \(L\) connecting the two points where the strict transform of \(C\) meets \(E\).  Using the usual Chern class identities for blowups, one readily checks that \(c_{1}(X)\) is not positive on the strict transform of \(L\) in \(X\).  The general argument of Appendix C builds on this idea using Mori theory.  A straightforward argument reduces an arbitrary hypothetical counterexample to this specific example and then rules it out analogously.
\begin{figure}[here]
\begin{center}
\includegraphics[totalheight=0.3\textheight]{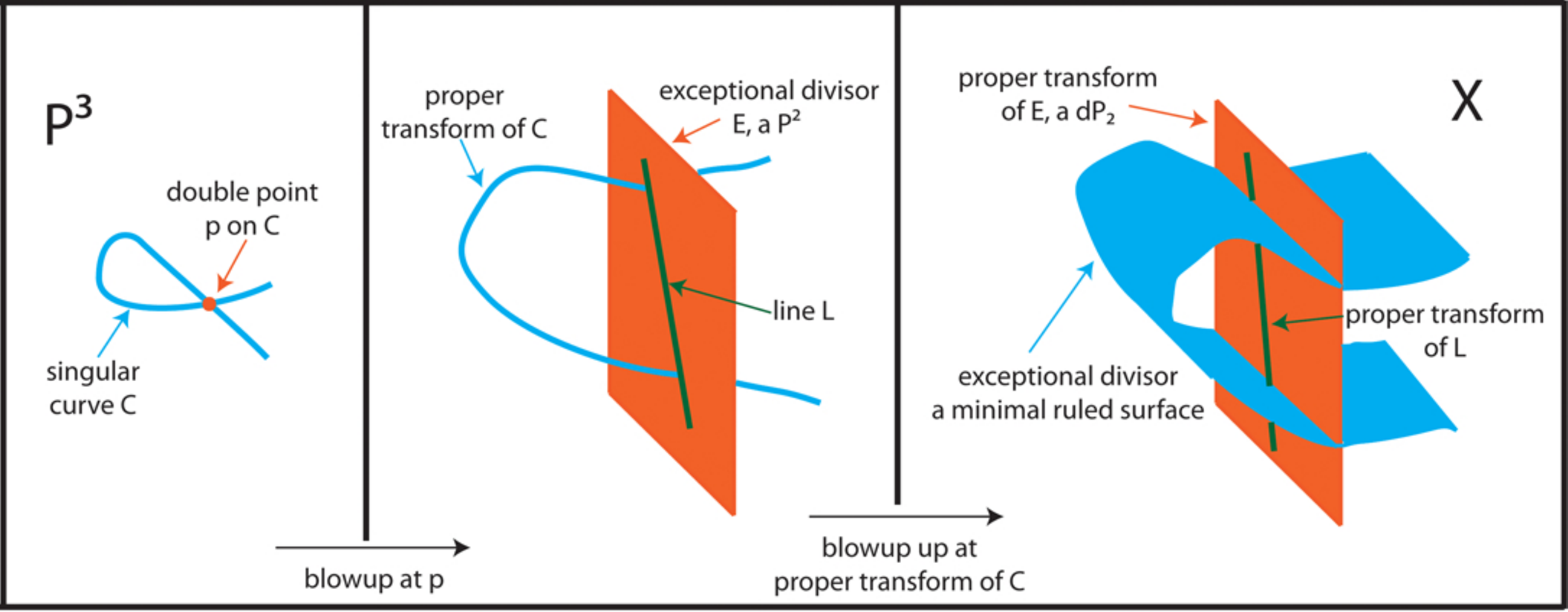}
\caption{
An attempt to create a Fano threefold with a \(dP_{2}\) which admits a decoupling limit by blowing up \(\mathbb{P}^{3}\) twice.  The construction fails because \(c_{1}(X)\) is not positive on the proper transform of the line \(L\).}
\label{fig:blowup}
\end{center}
\end{figure}
\paragraph{}
Thus we see that as candidate global completions of a local F-theory GUTs, Fano threefolds are ruled out.  In fact our arguments imply much more; these considerations are all local with respect to the compact part of the brane worldvolume \(S\) inside X.  It follows that if \(c_{1}(X)|_{S}\) is positive and \(S\) contains a sufficient number of mater curves to engineer the standard model, then \(S\) never admits a decoupling limit.  The fact that our arguments are local also makes them robust.  One cannot invalidate the conclusions by putting some horrible singularity of \(X\) far away from \(S\).  As long as the physics in a small neighborhood of the seven-brane is geometrically describable the analysis presented in this section goes through.  In fact, as demonstrated in Appendix C, even allowing singularities of \(S\) and  \(X\) only adds the possibility for \(S\) to be a the singular quadric cone mentioned in Section 2.  This result reinforces the conceptual link between asymptotic freedom of the gauge theory on the seven-brane and the existence of a limit where gravity decouples.  We have seen in Section 2 that curves \(C\subset X\) with \(c_{1}(X)\cdot C <0\) always lie in non-abelian singularities of the elliptic four-fold.  The fact that \(X\) cannot be Fano at \(S\) then tells us that we are tantalizingly close to being forced into the physically desired situation of non-abelian gauge theory simply by the existence of a decoupling limit.  
\begin{center}
\section{Constraints on Colliding Seven-Branes}
\label{obstruc}
\end{center}
\paragraph{}
Hopefully the analysis of Section \ref{kahlersec} has convinced the reader that Fano threefolds are bad candidate geometries for a global completion of a phenomenologically successful local F-theory GUT.  Thus in this section we will turn our attention to compactifications where the threefold \(X\) is not Fano.  The geometry in this case becomes more subtle, and the threefolds under investigation are now more varied and unclassified.  Nevertheless one can still make substantial progress by studying how the Ricci curvature of \(X\) behaves near our seven-brane.  As we have already seen, the Ricci curvature near the seven brane, i.e. the restriction of the first Chern class of \(X\) to \(S\), is intimately connected to the study of contractions of \(S\).  What we will now show is that it that \(c_{1}(X)\) also plays an essential role in limiting the local interactions and matter content allowed in any gauge theory supported on \(S\).
\subsection{Seven-Brane Tadpoles}
\paragraph{}
Intuitively it is clear that the curvature of \(X\) near \(S\) plays a role in constraining the allowed behavior of gauge theories on \(S\).  After all, matter is dictated by intersections of seven-branes, and since each seven brane produces a local curvature backreaction on the geometry, knowledge of \(c_{1}(X)|_{S}\) tells us something about what kinds of branes can meet \(S\).  The precise manifestation of this intuition is in the classification of degenerating elliptic fibers in the Weierstrass model.  Suppose for example that \(S\) supports a seven brane of \(SU(n)\) gauge group together with unspecified matter and Yukawa couplings.  Then from Table 1 we see that the vanishing loci of \(f\), \(g\), and \(\Delta\) take the form:
\begin{eqnarray}
\Delta& = & nS + D_{1}  \label{branevan} \\
g & = & D_{2} \label{bbranevan} \\
f & = & D_{3} \label{bbbranevan} 
\end{eqnarray}
Where in equations \((\ref{branevan})-(\ref{bbbranevan})\), the \(D_{i}\) denote effective divisors distinct from \(S\).  On the other hand since \(f\), \(g\), and \(\Delta\) are sections of \(4c_{1}(X)\), \(6c_{1}(X)\), and \(12c_{1}(X)\) equations  \((\ref{branevan})-(\ref{bbbranevan})\) encode properties about the canonical bundle of \(X\).  Viewed in this light, the equations \((\ref{branevan})-(\ref{bbbranevan})\) take the form of global seven-brane tadpole equations.  To see how they work in practice it is helpful to study equations \((\ref{branevan})-(\ref{bbbranevan})\) in Sen's IIB weak coupling limit \cite{Sen}.  For concreteness suppose we were studying F-theory on a Calabi-Yau fourfold whose base is \(\mathbb{P}^{3}\).  Let \(H\) denote a hyperplane in \(\mathbb{P}^{3}\).  The analog of equation \((\ref{branevan})\) then simply states that the total generalized seven brane charge as measured by the descriminant is \(-12K_{\mathbb{P}^{3}}=48H\).  We know from Sen's work is that in the weak coupling limit we can interpret this F-theory compactification as an orientifold compactification of IIB on a Calabi-Yau threefold \(B\).  Thus \(B\) should be a double cover of \(\mathbb{P}^{3}\).  The fact that \(B\) is Calabi-Yau tells us that the branch locus of the cover, where the orientifold planes reside, is in the class of \(-2K_{\mathbb{P}^{3}}=8H\).  Since \(O_{7}\) planes have \(D_{7}\) charge equal to \(-4\), we see that to cancel the \(D_{7}\) tadpole we must include in our compactification \(D_{7}\) branes whose total charge is \(32H\).  Now perturb slightly away from the weak coupling limit.  Sen tells us that each of the \(O_{7}\)'s resolve into a pair of mutually non-local \((p,q)\) seven branes, and hence the total generalized seven brane charge of the compactification is: 
\begin{equation}
\mathrm{7-Brane  \ Charge}= \underbrace{32H}_{contribution \ from \ D_{7}}+\underbrace{2\times8H}_{contribution \ from \ smeared \ out \ O_{7}}= 48H
\end{equation}
Exactly as required.  In the IIB limit these generalized tadpoles \((\ref{branevan})-(\ref{bbbranevan})\) will always be equivalent to the ordinary seven-brane tadpole condition, but away from the perturbative regime they force somewhat surprising relations on the configurations of seven-branes.
\paragraph{}
Another way to understand \((\ref{branevan})-(\ref{bbbranevan})\) is to note that in compactifications on lower dimensional manifolds they become much more familiar.  Indeed if we consider the foundational example of F-theory compactified on \(K3\) \cite{ftheory} then the tadpole equations tells us the well known fact that the total generalized seven-brane charge on the \(\mathbb{P}^{1}\) base of \(K3\) is \(-12K_{\mathbb{P}^{1}}=24\).    It is perhaps not a widely apprecciated fact that in compactifications on higher dimensional manifolds, tadpole cancelation becomes a much stronger requirement.  The point is that when we compactify F-theory on \(K3\) the generalized tadpole constraints are basically a relation among numbers.  On the other hand when we compactify F-theory on a Calabi-Yau fourfold, the seven-brane tadpoles are relations among cohomology classes of surfaces.  Given any such relation, we can intersect it with other cohomology classes to obtain new relations.  This last fact is particularly useful and powerful.  We can take equations \((\ref{branevan})-(\ref{bbbranevan})\) and restrict them to the brane surface \(S\) itself.  Then we obtain tadpole equations amongst cohomology classes on \(S\) that we can address in a local model, and which must be satisfied for any local seven-brane model to embed in string theory when gravity is turned back on.  
\paragraph{}
To make a systematic study of these restrictions, it is most convenient to prescribe the singularities of the elliptic fourfold using the Tate form of the equation for an elliptic curve:
\begin{equation}
y^{2}=x^{3}+a_{1}xy+a_{2}x^{2}+a_{3}y+a_{4}x+a_{6} \label{tate}
\end{equation}
Where now each \(a_{n}\) is a section of \(-nK_{X}\).  We will focus on the phenomenologically relevant example of an \(SU(5)\) gauge group on \(S\) though it should be clear that this method has more general application.  As with the Weierstrass form of the fourfold, the singularity type on a surface can be read off from the vanishing order of the \(a_{i}\).  Below we list only the groups relevant for us, a complete list can be found in \cite{geosing}.
\begin{center}
\begin{table}[h]
\begin{center}
\begin{spacing}{1.2}
\begin{tabular}{|c|c|c|c|c|c|c|c|l|}
\hline
Group & \(a_{1}\)& \(a_{2}\)& \(a_{3}\)& \(a_{4}\)& \(a_{6}\)& \(\Delta\) & Physical Meaning & Defining Equation\\
\hline
\(SU(5)\) & 0 & 1 & 2 & 3 & 5 & 5 & Gauge Fields on \(S\) & \(z=0\) \\
\hline
\(SU(6)\) & 0 & 1 & 3 & 3 & 6 & 6 & Charged Matter in \(\mathbf{5}\) & \(z=P=0\) \\
\hline
\(SO(10)\) & 1 & 1 & 2 & 3 & 5 & 7 &  Charged Matter in \(\mathbf{10}\) & \(z= b_{1}=0\) \\
\hline 
\(SO(12)\) & 1 & 1 & 3 & 3 & 5 & 8 & Yukawa Coupling \(\mathbf{\bar{5}}\) \(\mathbf{\bar{5}}\) \(\mathbf{10}\) & \(z = b_{1} =b_{3}=0\)\\
\hline 
\(E_{6}\) & 1 & 2 & 2 & 3 & 5 & 8 & Yukawa Coupling \(\mathbf{5}\) \(\mathbf{10}\) \(\mathbf{10}\)& \(z=b_{1}=b_{2}=0\)\\
\hline
\(SU(7)\) & 0 & 1 & 3 & 4 & 7 & 7 & Yukawa Coupling \(\mathbf{5}\) \(\mathbf{\bar{5}}\) \(\mathbf{1}\) & \(z=P=R=0\)  \\
 &  &  &  &  &  &  & &  \((b_{1},b_{3})\neq(0,0)\)\\
 \hline
\end{tabular}
\end{spacing}
\caption{Fourfold singularities for \(SU(5)\) gauge theories as specified by the vanishing orders of \(a_{i}\). }
\end{center}
\end{table}
\end{center}
Let \(z\) denote a local coordinate on \(X\) such that \(z=0\) locally defines \(S\).  Since we want \(SU(5)\) gauge symmetry on \(S\), in accordance with Table 2 we must choose:
\begin{equation}
a_{1}=b_{1} \hspace{.2in} a_{2}=zb_{2} \hspace{.2in} a_{3}=z^{2}b_{3} \hspace{.2in} a_{4}=z^{3}b_{4}\hspace{.2in} a_{6}=z^{5}b_{6} \label{avan}
\end{equation}
Where in equation \((\ref{avan})\) none of the \(b_{i}\) vanish identically on \(S\) where \(z=0\).  One readily computes that up to irrelevant constants, the discriminant of the equation \((\ref{tate})\) can be expanded in the following series in \(z\):
\begin{equation}
\Delta=z^{5}\left[b_{1}^{4}P+zb_{1}^{2}(8b_{2}P+b_{1}R)+\mathcal{O}(z^{2})\right] \label{deltasolve}
\end{equation}
And the quantities \(P\) and \(R\) are expressed in terms of the \(b_{i}\) as:
\begin{eqnarray}
P & = & b_{3}^{2}b_{2}-b_{1}b_{3}b_{4}+b_{1}^{2}b_{6} \label{Pdef} \\
R & = & 4b_{1}b_{2}b_{6}-b_{3}^{3}-b_{1}b_{4}^{2} \label{Rdef}
\end{eqnarray} 
Along curves in \(S\) the singularity type enhances to a rank one \(ADE\) extension of \(SU(5)\) and matter appears.  To deduce the precise charges one utilizes the Katz-Vafa procedure \cite{katz}.  For example, near the curve \(\Sigma_{\mathbf{5}}\) in \(S\) where the singularity enhances to \(SU(6)\) the local geometry is that of a single \(U(1)\) brane meeting the \(SU(5)\) brane \(S\) along \(\Sigma_{\mathbf{5}}\).   We can then view this theory as an \(SU(6)\) gauge theory which has been Higgsed to \(SU(5)\times U(1)\).  Since the \(SU(6)\) theory contains only adjoints, the matter content arising at the intersection of the seven-branes is then determined by reducing the adjoint of \(SU(6)\):
\begin{equation}
\mathbf{35}\rightarrow \mathbf{24}_{0}\oplus \mathbf{1}_{0}\oplus \mathbf{5}_{-1} \oplus \bar{\mathbf{5}}_{+1}
\end{equation}
Where the subscript refers to the \(U(1)\) charge.  Thus in terms of \(SU(5)\) representation content, the curve \(\Sigma_{\mathbf{5}}\) of \(SU(6)\) singularity enhancement hosts matter which transforms in the fundamental and antifundamental of \(SU(5)\).  Similarly, along a curve \(\Sigma_{\mathbf{10}}\) the singularity increases to \(SO(10)\) and matter in the \(\mathbf{10}\) and \(\mathbf{\overline{10}}\) appears.  
\paragraph{}
It is significant that the locations of the matter curves are completely fixed by the local behavior of the \(a_{i}\) near \(S\) in equation \((\ref{avan})\).  Since these are all sections of powers of \(-K_{X}\) the adjunction formula then connects the matter curves with the normal bundle of \(S\) in \(X\) and the canonical bundle of \(S\) itself.  To see this explicitly observe that the locus of \(SO(10)\) enhancement is exactly defined by \(b_{1}=0\).  On the other hand, we see that from \((\ref{avan})\) that \(b_{1}\) is a section of \(-K_{X}\).  Thus as cohomology classes:
\begin{equation}
\Sigma_{\mathbf{10}}=c_{1}(X)|_{S}=c_{1}(S)+c_{1}(N_{S/X}) \label{tensolve}
\end{equation}
Similarly, the \(SU(6)\) locus is determined by \(P=0\).  Recalling that \(\Delta\) is in the class of \(12c_{1}(X)\), homogeneity of equation \((\ref{deltasolve})\) implies: 
\begin{equation}
\Sigma_{\mathbf{5}}=8c_{1}(X)|_{S}-5S\cap S =8c_{1}(S)+3c_{1}(N_{S/X}) \label{fivesolve}
\end{equation} 
Where in \((\ref{fivesolve})\) we have used the fact that the self-intersection of \(S\) represents the Chern class of the normal bundle of \(S\) in \(X\).  A priori, we do not know exactly what the normal bundle of \(S\) is, however we see that we can eliminate the normal bundle from equations \((\ref{tensolve})-(\ref{fivesolve})\) to obtain a constraint:
\begin{equation}
3\Sigma_{\mathbf{10}}-\Sigma_{\mathbf{5}}+5c_{1}(S)=0 \label{constraint1}
\end{equation}
In other words, local seven-brane models have significantly less freedom then one might expect.  It is not possible to specify arbitrarily the matter curves on the seven-brane.  Once the surface \(S\) and class of the \(\mathbf{10}\) curve are specified, the class of the \(\mathbf{5}\) curve is determined by equation \((\ref{constraint1})\).  Some comments on equations \((\ref{tensolve})-(\ref{constraint1})\) are in order:
\begin{enumerate}
\item The derivation of these equations following this method was first performed in \cite{Donagi}.  There it was observed following \cite{Sadov} that the constraint \((\ref{constraint1})\) has an interpretation in the IIB weak coupling limit in terms of the Green-Schwarz cancellation of the mixed gauge gravitational anomalies on the seven-brane gauge theory.  The idea is that we can view each of the matter curves as a six-dimensional charged defect in the twisted gauge theory on the seven-brane \(S\), and so under a general gauge and Lorentz transformation the action acquires a variation localized on the matter curves which must be cancelled appropriately by the variation due to bulk fields.
\item Equations \((\ref{tensolve})-(\ref{constraint1})\) allow us to resolve the puzzle about the mismatch between the holomorphic normal deformations of \(S\) and the light adjoints arising from the dimensional reduction of \(\phi\).  In Section 3 we attributed this mismatch to the fact that some of the normal deformations of \(S\) cannot be consistently extended to the remaining branes in the geometry.  In order for this interpretation to make sense we would expect that in the seven-brane theory with no matter, \(N_{S/X}\) matches with \(K_{S}\), the latter being where the field \(\phi\) is valued.  Examining equations \((\ref{tensolve})\) we see that when \(\Sigma_{\mathbf{10}}\) vanishes we indeed have the equality \(K_{S}=N_{S/X}\) implying that the ambient geometry is locally Calabi-Yau.  This is as one might expect from perturbative IIB considerations; the \(\mathbf{10}\) curve is the locus of intersection of the seven-brane \(S\) with any orientifold planes so a necessary condition for \(X\) to be locally Calabi-Yau near \(S\) is the absence of any \(\mathbf{10}\) curves.  Curiously, if one further requires that the theory contain no \(\mathbf{5}\) curves then the unique solution to \((\ref{tensolve})-(\ref{constraint1})\) is \(c_{1}(S)=N_{S/X}=0\) so the only seven-brane theory without matter which can be consistently coupled to gravity has \(S=K3\) in an ambient Ricci flat geometry. 
\paragraph{}
More generally the constraint \((\ref{tensolve})\) shows that the light adjoints in the theory descending from the \(\phi\) field are in one-to-one correspondence with the subset of holomorphic normal deformations which vanish along the curve \(\Sigma_{\mathbf{10}}\).   In the IIB limit this reflects the simple fact that at \(\Sigma_{\mathbf{10}}\), \(S\) meets its mirror image at the orientifold plane, and any allowed motion of the seven-brane configuration must respect this fact.  
\end{enumerate}
\paragraph{}
Using the techniques demonstrated thus far we can extend the constraints \((\ref{tensolve})-(\ref{constraint1})\) to a method for counting the Yukawa couplings in the geometry.  As discussed in the introduction, Yukawa coupling are generated when matter curves intersect and hence the singularity type enhances by a rank-two extension of \(SU(5)\).  The relevant rank two enhancements together with their associated physical interpretation are cataloged in Table 2.  We will denote by \(p(G)\) for \(G= SU(7), SO(12), E_{6}\) the number of points in \(S\) generating each of the indicated Yukawas.  Examining Table 2 and equations \((\ref{avan})-(\ref{deltasolve})\) and using homogeneity of the discriminant as above we find:
\begin{eqnarray}
p(SO(12))  & = & b_{1} \cap b_{3} =  \left(\phantom{\int}\hspace{-.17in}3c_{1}(X)\cdot c_{1}(X) -2c_{1}(X)\cdot S\right) \cdot S   \\
p(E_{6})  & = & b_{1} \cap b_{2} =   \left(\phantom{\int}\hspace{-.17in}2c_{1}(X)\cdot c_{1}(X) - c_{1}(X)\cdot S\right ) \cdot S \\
p(SU(7))  & = & P \cap R -2b_{1} \cap b_{3}  =  \left(\phantom{\int}\hspace{-.17in} 66c_{1}(X)\cdot c_{1}(X) -89c_{1}(X)\cdot S+30S\cdot S \right)\cdot S \label{doublecount}
\end{eqnarray} 
The only small subtlety in deriving these formulas occurs in the left-hand-side of equation \((\ref{doublecount})\).  Points where \(b_{1}\) meets \(b_{3}\) define \(SO(12)\) points and also occur in the intersection \(P\cap R\); in other words an \(SO(12)\) point is a special case of an \(SU(7)\) point.\footnote{I would like to thank Mboyo Esole for patiently explaining this to me.}  To avoid overcounting the \(SU(7)\) Yukawa points we must then subtract these \(b_{1}\cap b_{3}\) points from the \(P\cap R\) points.  The reason for the factor of two in the subtraction is then that points in \(P\cap R\) where \(b_{1}=b_{3}=0\) are intersections with multiplicity two, as can easily be seen from the defining equations \((\ref{Pdef})-(\ref{Rdef})\).
\paragraph{}
Now, because of equation \((\ref{tensolve})\) these polynomials in \(c_{1}(X)\) and \(S\) can be reduced to \emph{purely local data} about the seven-brane \(S\) and the \(\mathbf{10}\) curve \(\Sigma_{\mathbf{10}}\) with result:
 \begin{eqnarray}
p(SO(12))  & = &  \Sigma_{\mathbf{10}}\cdot \Sigma_{\mathbf{10}}+2 \Sigma_{\mathbf{10}} \cdot c_{1}(S) \label{so12} \\
p(E_{6}) &  = & \Sigma_{\mathbf{10}}\cdot \Sigma_{\mathbf{10}}+\Sigma_{\mathbf{10}}\cdot c_{1}(S) \label{e6} \\
p(SU(7))  & = & 7 \Sigma_{\mathbf{10}}\cdot \Sigma_{\mathbf{10}}+29\Sigma_{\mathbf{10}}\cdot c_{1}(S)+30 c_{1}(S)\cdot c_{1}(S) \label{su7} 
\end{eqnarray} 
Again we see that the local freedom in a seven-brane model is less than expected.  Once \(S\) and \(\Sigma_{\mathbf{10}}\) are chosen, the number of Yukawa couplings of each type are determined.  In fact, one linear combination of these Yukawas is even independent of the matter curves and sensitive only to the brane worldvolume \(S\):
\begin{equation}
p(SU(7))+15p(E_{6})-22p(SO(12))=30c_{1}(S)\cdot c_{1}(S) \label{yuksum}
\end{equation}
It is unclear to us what, if any, the precise gauge theory interpretation of these constraints are.  The fact that they are derived analogously to the anomaly equation \((\ref{constraint1})\) suggests a relation to anomaly cancelation and the Green-Schwarz mechanism, this time constraining the number and kind of four-dimensional defects in the compactified seven-brane gauge theory.  In any case, the method of restricting the seven-brane tadpoles to a seven-brane and studying their intersection provides a simple derivation of \((\ref{su7})-(\ref{e6})\), and it is easy to check that these constraints are satisfied in all known globally consistent examples \cite{Grimm2} \cite{Donagi} \cite{Mar}. 
\paragraph{}
It is important to understand the implications of the anomaly equation \((\ref{constraint1})\) and Yukawa constraints \((\ref{su7})-(\ref{e6})\) for the local models constructed by Heckman, Vafa, and collaborators.  Taking as a representative example \cite{BHVII}, one finds a presentation of matter curves and a choice of surface \(S\).  If one interprets their construction in the strictest sense as a claim that there exist only those matter curves and nothing more, then their models are obstructed from UV completion by the constraints derived in this section.  However, a more reasonable interpretation of their work is that the matter curves enumerated represent only a proper subset of the complete brane intersection locus.  Indeed while for generic seven-brane intersections the \(\mathbf{10}\) curve \(\Sigma_{\mathbf{10}}\) is a single connected curve, it is certainly possible that for a suitably prescribed intersection the \(\mathbf{10}\) curve splits into, say, two pieces \(\Sigma_{\mathbf{10}}=C_{1}+C_{2}\) where \(C_{1}\) contains the piece of the \(\mathbf{10}\) curve appearing in a Heckman-Vafa model and \(C_{2}\) is chosen to satisfy the anomaly equation \((\ref{constraint1})\).  To demonstrate consistency of their models it is thus necessary to exhibit an explicit splitting of the brane intersection locus which does not modify the phenomenology.  We will refer to this problem as a \emph{factorization problem}: one must factorize the brane intersection locus in order to satisfy the constraints.\footnote{The existence of this problem, if not its underlying cause was first recognized in \cite{Mar}.}  It is clear that this is a feature of brane constructions which can and should be addressed in a purely local model.  Although we will not discuss this problem in detail, the restrictions on models with decoupling limits derived in Section 4.2 will likely prove useful in attacking this issue.
\paragraph{}
It may be possible to solve the factorization problem while leaving no residue of its existence in four-dimensions.  The charged four-dimensional matter fields in the theory are the zero-modes of the fields on the matter curves, so in the notation of the previous paragraph it could be that the curve \(C_{2}\) supports no zero-modes and therefore does not effect the four-dimensional action.  There is at least one significant question to be addresed in solving the factorization problem in this way: 
\begin{itemize}
%\item \emph{Why do the additional required curves have no zero-modes?}  Brane fluxes give us some ability to choose the net chiral spectrum, but in general there will still be vector-like pairs of zero-modes.  In a generic effective field theory one typically does not worry about the existence of vector-like pairs since their mass is not protected by any symmetry.  However, seven-brane GUT models are not generic effective field theories and one must exhibit an actual mechanism for lifting these vector-like pairs from the spectrum.  In the absence of such a mechanism these fields will remain massless until supersymmetry is broken and in general effect the phenomenology.  This is a particularly important issue for GUT models as the running of non-abelian gauge couplings is extremely sensitive to the existence of additional charged matter.
\item \emph{What stabilizes the factorization?}  This is clearly a subissue of the general problem of moduli stabilization.  It seems that some degree of factorization will be required purely by phenomenology.  For example, it is reasonable to surmise that the \(\mathbf{5}\) curve must be split into a least three pieces:
\begin{equation}
\Sigma_{\mathbf{5}}=C_{\mathbf{5}_{H}}+C_{\bar{\mathbf{5}}_{H}}+C_{\bar{\mathbf{5}}_{M}}+\cdots \label{fivesplit}
\end{equation} 
Where in equation \((\ref{fivesplit})\), \(C_{\mathbf{5}_{H}}\) denotes a curve supporting the \(\mathbf{5}\) Higgs field, \(C_{\bar{\mathbf{5}}_{H}}\) denotes a curve supporting the \(\mathbf{\bar{5}}\) Higgs field, and \(C_{\bar{\mathbf{5}}_{M}}\) a curve supporting the \(\bar{\mathbf{5}}\) matter field.  It is unclear whether stabilizing an additional factorization beyond that required by phenomenology will be any more challenging then the general problem of moduli stabilization faced by any viable model.  Returning to the particular constructions of Heckman, Vafa, et. al., it is natural to expect that the enhanced symmetry structures in \cite{M1} \cite{M3} will play a fundamental role in stabilizing the required factorization in their models.
\end{itemize}
\paragraph{}
In the remainder of the paper when we make assertions about the phenomenological implications of our results we will have in mind models that do not face a factorization problem beyond that demanded by phenomenology.  We will take as our working definition of a minimal generic F-theory GUT a model where there exists a single connected \(\mathbf{10}\) curve and a \(\mathbf{5}\) curve split into three pieces corresponding to the three MSSM matter curves in equation \((\ref{fivesplit})\), \emph{and nothing else}.  These curves and their intersections will then be chosen in order to satisfy the constraints \((\ref{constraint1})\), \((\ref{su7})\), \((\ref{so12})\), \((\ref{e6})\).  Further, we will assume that the points of intersection of these curves are uncorrelated and constrained only by basic phenomenological requirements of, for example, matter parity.\footnote{Supersymmetry breaking may require additional fields.  For example in a gauge mediated scenario one could envision further factorizing say the \(\mathbf{5}\) curve to include an additional piece supporting a vector like pair of messenger fields \cite{M5} \cite{Mar2}.  Because we must require messenger matter couplings to vanish this additional curve will not effect the assertions made in the remainder of the paper and can safely be ignored.}  Our purpose in these assumptions is not to claim that these models are preferred.  On the contrary, we will see in Section 4.2 that for purely local models these assumptions are in fact a bit too strong.  We use these models as examples because for these, the constraints derived in this section are the most powerful. 
\paragraph{}
For the class of F-Theory GUTs defined above, there is a simple consequence of the Yukawa constraint \((\ref{e6})\) that is worth mentioning.  To this end we must first recall the four-dimensional meaning of the number of Yukawa points \cite{BHVII} \cite{M4}.   Along a matter curve, say \(\Sigma_{\mathbf{10}}\), resides a six-dimensional defect theory coupled to the seven-brane gauge theory on \(S\).  The \(SU(5)\) representation on the matter curve is determined by the Katz-Vafa Higgsing procedure reviewed above.  In particular since the representation always results from breaking an adjoint, the six-dimensional theory on the matter curve is vector-like.  To obtain a four-dimensional chiral spectrum we now switch on a brane flux on \(S\) and dimensionally reduce.  Say for example we find \(k\) chiral zero modes, and let their wavefunctions on \(\Sigma_{\mathbf{10}}\) be \(\psi_{1}(w), \cdots \psi_{k}(w)\), where \(w\) denotes a local coordinate on \(\Sigma_{\mathbf{10}}\), and \(w=0\) is a point in \(S\) where a Yukawa coupling is generated.  The zero modes can be organized according to their vanishing order at \(w=0\):
\begin{equation}
\psi_{j}(w)\sim w^{j-1} \label{wv}
\end{equation}
At the point \(w=0\), three matter curves meet and to leading order the Yukawa coupling for the zero-modes involved is simply given by the product of the three wavefunctions at the Yukawa point \cite{BHVI} \cite{Wijn}.  According to \((\ref{wv})\) all but one of the zero modes vanishes at this point so we see that a single Yukawa point in \(S\) leads to a rank one matrix of four-dimensional Yukawas for the zero modes on \(\Sigma_{\mathbf{10}}\).  More generally when there are multiple points generating Yukawas for the zero modes on \(\Sigma_{\mathbf{10}}\) the basis with the simple behavior \((\ref{wv})\) will be different for each point, so the previous argument implies that the number of Yukawa points for the matter curve \(\Sigma_{\mathbf{10}}\) is the rank of the four-dimensional Yukawa matrix for the zero modes on \(\Sigma_{\mathbf{10}}\).\footnote{Obviously the rank is bounded above by the number of zero modes, so once the number of Yukawa points exceeds the number of zero modes we simply have maximal rank.}
\paragraph{}
Now examine \((\ref{e6})\).  For generic brane moduli \(\Sigma_{\mathbf{10}}\) is a single connected curve and supports all three standard model generations of \(\mathbf{10}\).  We can apply the genus formula:
\begin{equation}
p(E_{6})=\left(\Sigma_{\mathbf{10}}\cdot \Sigma_{\mathbf{10}}-\Sigma_{\mathbf{10}}\cdot c_{1}(S)\right)+2\Sigma_{10}\cdot c_{1}(S)= 2g(\Sigma_{10})-2+2\Sigma_{10}\cdot c_{1}(S) \label{evene}
\end{equation}
In particular, the right-hand-side of \((\ref{evene})\) is even.  In accordance with the arguments above we conclude that the Yukawa matrix for the \(E_{6}\) coupling \(\mathbf{5} \ \mathbf{10} \ \mathbf{10}\) has even rank.  On reduction to standard model gauge group \(SU(3)\times SU(2)\times U(1)\) this coupling is responsible for the mass of up-type quarks, and hence to leading order has rank one amongst the observed standard model spectrum.  The minimal solution of \((\ref{evene})\) consistent with low-energy data is then not three \(\mathbf{10}\) zero modes with a rank one Yukawa, but rather four \(\mathbf{10}\) zero modes with a rank two Yukawa.  In other words:
\emph{For generic brane moduli which can accommodate the standard model, F-theory GUTs predict the existence of additional \emph{\textbf{10}}'s.  }  In keeping with the genericity assumption one might expect that the two eigenvalues of this matrix are roughly of the same order, in which case these additional quarks should not be too much heavier than the top quark.  This is certainly possible while staying in experimental bounds.  For example, if one adds a complete fourth generation to the standard model the bound on the mass of the up type quark \(t'\) is \(m_{t'}> 256\)GeV \cite{genfour}.\footnote{In order to avoid constraints from electroweak precision observables it is necessary that there be a minor mass hierarchy between the new bottom type quark and the up type.  All this and more is reviewed in the cited reference.}  
\subsection{Decoupling Limits and Examples}
\label{exotic}
\paragraph{} In section 4.1 we derived a number of global constraints on any local F-theory \(SU(5)\) GUT.  These are a priori constraints on the form of local singularities of any compact elliptically fibered Calabi-Yau fourfold with section and are valid independent of the existence of a decoupling limit.  If we now assume further that our seven-brane gauge theory can be consistently decoupled from gravity, then the anomaly and Yukawa constraints acquire new power due to the fact that we now have independent knowledge of the normal bundle of \(S\).  For example, consider equation \((\ref{tensolve})\) as a relation amongst line bundles on \(S\):
\begin{equation}
N_{S/X}-\Sigma_{\mathbf{10}}=K_{S} \label{ruled2}
\end{equation}
If \(S\) admits a decoupling limit then by negativity of contraction, no positive power of \(N_{S/X}\) admits a holomorphic section, and hence by \((\ref{ruled2})\) no positive power of the canonical bundle admits any section.  As mentioned in Section 3 this implies that \(S\) is a ruled surface, a \(\mathbb{P}^{1}\) fibration over any smooth complex curve, blown up at points an arbitrary number of times.
\paragraph{}
Mathematically there is absolutely no constraint on the genus \(g\) of the base of this ruled surface.  But for \(g>0\) the surface \(S\) has non-trivial holomorphic one-forms \(h^{1,0}(S)=g\).  In the four-dimensional effective theory these one-forms give rise to adjoint chiral superfields descending from the reduction of the gauge field on the seven-brane.  The difficulty for \(g>0\) seems to be that there is no way to generate a holomorphic mass for these adjoints, at least if we utilize only the seven-brane gauge theory on \(S\) \cite{BHVII} \cite{bourjaily} \cite{Wijn}.  The adjoints descending from holomorphic one-forms 
\begin{wrapfigure}{r}{.45\textwidth}
\fbox{
\begin{minipage}{2.75in}
\begin{center}
\includegraphics[totalheight=0.7\textwidth]{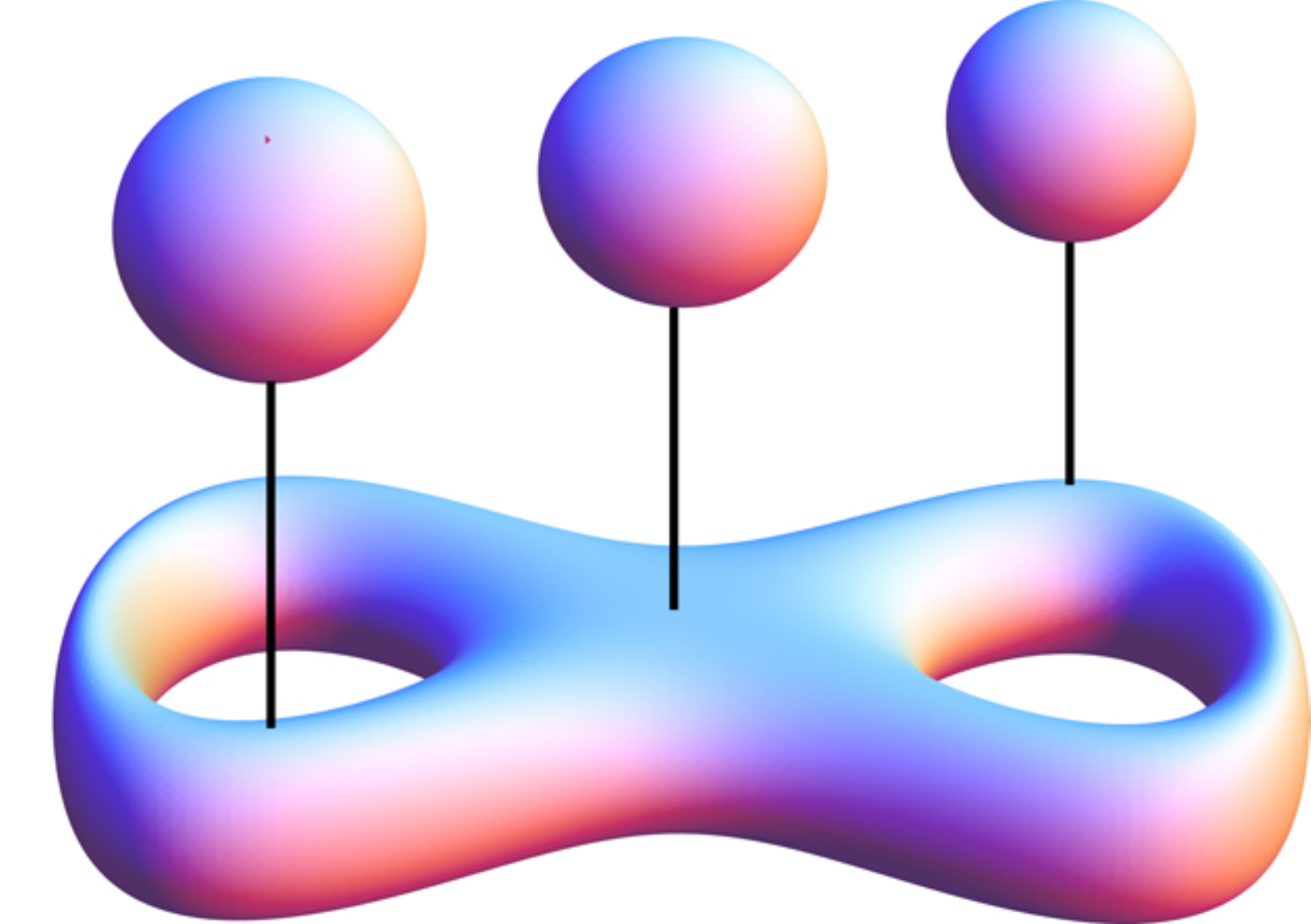}
\end{center}
\caption{An example of a ruled surface \(S\) is a \(\mathbb{P}^{1}\) fibration over a curve of genus two.  A seven-brane gauge theory compactified on \(S\) would have a pair of phenomenologically undesirable light adjoint scalars.}
\end{minipage}
}
\end{wrapfigure}
have couplings only to vector-like pairs of zero modes and so in the absence of supersymmetry breaking remain massless.  In practice this means that we must exclude these adjoints from the spectrum.  Indeed even a single \(SU(3)\) adjoint of the QCD color gauge group that persists to the weak scale is enough to force \(\alpha_{strong}\) to reach a Landau pole before the GUT scale.  Thus from now on we will impose as a second phenomenological constraint \(h^{1,0}(S)=g=0\).  The surfaces \(S\) which remain are then the so-called \emph{rational surfaces} which can be obtained by blowups of the Hirzebruch surface \(\mathbb{F}_{n}\).  We will use the notation \(\mathbb{B}_{k}\mathbb{F}_{n}\) to indicate a blowup at \(k\) points, and for the remainder of the paper \(S\) will always denote such a surface.  Our requirement discussed in Section 3 of a sufficient number of matter curves then tells us that \(k\geq1\).  A significant feature of all of these surfaces is that they are all simply connected and hence admit no Wilson lines.  In particular, this means that the only known way to Higgs the GUT group is to use brane flux \cite{Donagi}.  
\paragraph{}
We have seen that there are a number of different ways in which a seven-brane might admit a decoupling limit.  By far the simplest however is that the compact part of the brane worldvolume should undergo an elementary contraction to a point.  In this case one can be sure that no additional geometric scales from \(X\) enter in the gauge theory on the seven-brane.  Using the techniques developed thus far it is now easy to show no \(\mathbb{B}_{k}\mathbb{F}_{n}\) which carries an \(SU(5)\) brane admits such a decoupling limit.  To show this, all we need to do is run the argument in Section 3 used to rule out Fanos in reverse.  We know that \(S\) has a Mori cone of curves spanned by rational curves \(\Gamma_{i}\) with negative normal bundle, and further by Grauert \(c_{1}(N_{S/X})\) is negative on all of the \(\Gamma_{i}\).  We again apply the genus formula:
\begin{equation}
0=g(\Gamma_{i})=\frac{1}{2}(\Gamma_{i}-c_{1}(S))\cdot \Gamma_{i}+1=\frac{1}{2}(c_{1}(N_{S/X})-c_{1}(X))\cdot \Gamma_{i} +\frac{\Gamma_{i}\cdot \Gamma_{i}}{2}+1 \label{genusf1}
\end{equation}
And this time we conclude that \(-c_{1}(X)\cdot \Gamma_{i} \geq 0\).  Since any curve in \(S\) is a positive integral sum of the \(\Gamma_{i}\) we conclude that in fact \(-c_{1}(X) \geq 0\) on all of \(S\).  If \(c_{1}(X)\) vanishes on \(S\) then by equation \((\ref{tensolve})\) there is no \(\mathbf{10}\) curve, a phenomenological disaster.  Thus we must have \(-c_{1}(X)>0\) on at least one curve, or what is equivalent, \(-c_{1}(X)\) must be positive on at least one curve \(C\subset X\) with \(C\cdot C \geq0\).  Applying the same logic used in Appendix B to deduce that negativity of contraction implies no holomorphic normal deformations, we conclude that every section of the bundle \(-K_{X}\) vanishes identically on \(S\).  In particular in the notation of Table 2, \(a_{1}\) must vanish so such a surface never supports an \(SU(5)\) type brane.  In the language of IIB, there must be an orientifold plane directly on top of our seven-brane.
\paragraph{}
We can gather more information about seven-brane models with decoupling limits by incorporating the considerations of Section 4.1.  As a prelude to this, it is useful to review the cohomology of the candidate seven-brane surfaces \(\mathbb{B}_{k}\mathbb{F}_{n}\).   The Hodge diamond of these surfaces has the following shape:
\begin{eqnarray}
\begin{tabular}{cccc}
\multicolumn{4}{c}{\(h^{2,2}(S)\)}  \\
\multicolumn{4}{c}{\(h^{2,1}(S)\) \hspace{.15in} \(h^{1,2}(S)\)} \\
\multicolumn{4}{c}{\(h^{2,0}(S)\) \hspace{.15in} \(h^{1,1}(S)\) \hspace{.15in} \(h^{0,2}(S)\)} \\
\multicolumn{4}{c}{\(h^{1,0}(S)\) \hspace{.15in} \(h^{0,1}(S)\)} \\
\multicolumn{4}{c}{\(h^{0,0}(S)\)}  
\end{tabular}
 & = &  
\begin{tabular}{cccc}
\multicolumn{4}{c}{1}  \\
\multicolumn{4}{c}{0 \hspace{.175in} 0} \\
\multicolumn{4}{c}{0 \hspace{.15in} \(2+k\) \hspace{.15in} 0} \\
\multicolumn{4}{c}{0 \hspace{.175in} 0} \\ 
\multicolumn{4}{c}{1} 
\end{tabular} 
\label{hodged2}
\end{eqnarray}
\(H^{1,1}(\mathbb{B}_{k}\mathbb{F}_{n})\) is generated by \(2+k\) cohomology classes, \(B\), \(F\), and \(E_{i}\) for \(i=1, \cdots , k\).  Each of these classes is represented in \(\mathbb{B}_{k}\mathbb{F}_{n}\) by a rational curve.  The intersections of these classes are given by:
\begin{equation}
B\cdot B = -n \hspace{.5in} B\cdot F =1 \hspace{.5in} E_{i}\cdot E_{j}=-\delta_{ij} \hspace{.5in} F\cdot F = F\cdot E_{i}= B \cdot E_{i}=0 \label{intrela}
\end{equation}
Finally, the first Chern class of \(\mathbb{B}_{k}\mathbb{F}_{n}\) is then given by:
\begin{equation}
c_{1}(S)= 2B+(n+2)F-\sum_{i=1}^{k}E_{i} \label{cherns}
\end{equation}
The constraints derived in Section 4.1 suggest that to refine our understanding of seven-brane models with decoupling limits we should constrain the \(\mathbf{10}\) curve \(\Sigma_{\mathbf{10}}\).  One way to do this is to apply index theory to the normal bundle \(N_{S/X}\).  The existence of a decoupling limit implies in particular that \(S\) is rigid so \(h^{0}(S,N_{S/X})=0\).  Similarly by Serre duality, adjunction, and our derivation \((\ref{tensolve})\) we have:
\begin{equation}
h^{2}(S,N_{S/X})=h^{0}(S,K_{S}-N_{S/X})=h^{0}(S,K_{X}|_{S})=h^{0}(S,-\Sigma_{\mathbf{10}})=0 \label{indexmadness}
\end{equation}
Where the last equality in \((\ref{indexmadness})\) follows from the elementary fact that \(\Sigma_{\mathbf{10}}\) is an effective divisor in \(S\).  In particular, we deduce that the holomorphic Euler characteristic \(\chi_{\hspace{-.05in}\phantom{a}_{Hol}}(S,N_{S/X})\) satisfies the inequality:
\begin{equation}
\chi_{\hspace{-.05in}\phantom{a}_{Hol}}(S,N_{S/X})=h^{0}(S,N_{S/X})-h^{1}(S,N_{S/X})+h^{2}(S,N_{S/X})=-h^{1}(S,N_{S/X})\leq0 \label{ineq1}
\end{equation}
On the other hand the quantity \(\chi_{\hspace{-.05in}\phantom{a}_{Hol}}(S,N_{S/X})\) can independently be computed by an application of the index theorem:
\begin{equation}
\chi_{\hspace{-.05in}\phantom{a}_{Hol}}(S,N_{S/X})= \int_{S}Ch(N_{S/X})Td(S)=1+\frac{c_{1}(N_{S/X})\cdot c_{1}(S)}{2}+\frac{c_{1}(N_{S/X})\cdot c_{1}(N_{S/X})}{2} \label{indexm2}
\end{equation}
Where in \((\ref{indexm2})\) we have used the intersection ring of \(\mathbb{B}_{k}\mathbb{F}_{n}\) to simplify the right-hand-side.  Now we eliminate \(c_{1}(N_{S/X})\) in favor of \(\Sigma_{\mathbf{10}}\) using \((\ref{tensolve})\).  Combining \((\ref{indexm2})\) with the inequality \((\ref{ineq1})\) we obtain:
\begin{equation}
\frac{\Sigma_{\mathbf{10}}\cdot \Sigma_{\mathbf{10}}-\Sigma_{\mathbf{10}}\cdot c_{1}(S)}{2}+1 \leq0 \label{ineq2}
\end{equation}
The result \((\ref{ineq2})\) provides useful information about any seven-brane model with a decoupling limit.  In the simplest class of such models \(\Sigma_{\mathbf{10}}\), is single connected curve in which case the left-hand-side of \((\ref{ineq2})\) is simply the genus of this curve.  Since the genus of a curve is never negative for these examples, equation \((\ref{ineq2})\) states that \(\Sigma_{\mathbf{10}}\) is a smooth \(\mathbb{P}^{1}\).  More generally in models where the \(\mathbf{10}\) curve is factorized into a number of pieces, equation \((\ref{ineq2})\) significantly constrains the the intersections of the components.
\paragraph{}
A second general result with interesting implications concerns the structure of Yukawa couplings.  For the phenomenological success of our model, we require non-vanishing up and down type Yukawa matrices, so \(p(E_{6})\) and \(p(SO(12))\) must be positive.  On the other hand, we have seen in \((\ref{e6})\) that the number of \(E_{6}\) Yukawa coupling points can be expressed in terms of the \(\mathbf{10}\) curve.  Thus:
\begin{equation}
p(E_{6})=\Sigma_{\mathbf{10}}\cdot \Sigma_{\mathbf{10}}+c_{1}(S)\cdot \Sigma_{\mathbf{10}} >0 \label{ineq3}
\end{equation}
On combining the two inequalities \((\ref{ineq2})-(\ref{ineq3})\) we then have:
\begin{equation}
c_{1}(S)\cdot \Sigma_{\mathbf{10}}=p(SO(12))-p(E_{6})>1 \label{nondecoup}
\end{equation}
Furthermore, by combining \((\ref{nondecoup})\) with the Yukawa sum relation \((\ref{yuksum})\) on \(S\cong \mathbb{B}_{k}\mathbb{F}_{n}\)  we find:
\begin{equation}
p(SU(7)) > 262-30k \label{neutrino}
\end{equation}
Let us discuss the latter of these inequalities first.  The \(SU(7)\) type Yukawa points give rise the the interaction \(\bar{\mathbf{5}} \ \mathbf{5} \ \mathbf{1}\), where \(\mathbf{1}\) denotes standard model singlets localized on matter curves on branes transverse to \(S\).  A simple candidate interpretation of these singlets is that they are right-handed neutrinos which acquire Majorana masses from dynamics not confined to \(S\).  Integrating out these heavy fields from the four-dimensional effective action, we then find a neutrino mixing matrix whose structure is determined by the SU(7) Yukawas.  At least for small \(k\), \((\ref{neutrino})\) implies that there are a large number of uncorrelated points where this Yukawa is generated so generically we would expect a completely anarchic structure.  To be concrete, the del Pezzo models studied in the recent F-theory literature all have \(1\leq k\leq 7\) in which case a typical number of \(SU(7)\) points is in the hundreds.
\paragraph{}
The implications of \((\ref{nondecoup})\) are significantly more dramatic.  The inequality \((\ref{nondecoup})\) implies that in models with a decoupling limit there is necessarily a mismatch between the number of \(SO(12)\) couplings and the number of \(E_{6}\) couplings.  It follows from our analysis in Section 4.1 that for a generic minimal F-theory GUT we expect a mismatch in rank between the up and down type Yukawa matrices.  This is a phenomenological disaster.  To avoid this conclusion we must break the genericity or minimality assumption in some way.  One particularly natural idea first proposed in \cite{M1} and \cite{M3} for different reasons is to correlate the points in \(S\) where the Yukawa couplings are generated by bringing them close together.  In this case, the bases of zero modes with the nice behavior \((\ref{wv})\) at the Yukawa points are related because the interaction points are nearby, and hence our assertion that the number of Yukawa points is the rank of the corresponding four-dimensional Yukawa matrix is violated.  In general, one would expect that to stabilize this additional structure would require an additional symmetry and there is an obvious candidate: \emph{consider a Yukawa coupling where \(SU(5)\) enhances by more than a rank two extension.}  From the point of view of the \(SU(5)\) model with the generic Yukawa points, this means that we have put several interactions directly on top of each other.  This is an intriguing possibility, and though beyond the scope of this paper, it would be interesting to understand the a priori constraints on such exotic point-like singularities analogous to \((\ref{su7})-(\ref{e6})\).
\paragraph{}
After this brief general overview of decoupling seven-brane models we now turn to more specific scenarios indexed by the local model of \(X\) near \(S\).  The no-go result presented in this section implies that our seven-brane either decouples from gravity by undergoing an elementary contraction to a curve, or a non-elementary contraction to a point or a curve.  In what follows we will highlight some interesting gross features of theses scenarios.  Our discussion is rather brief, and the physics remains to be understood in detail.
\subsubsection{Elementary Contraction to a Curve}
\paragraph{}
In this case, the surface \(S\) shrinks to a curve and no other surface which meets \(S\) shrinks in the decoupling limit.  A globally complete example of this type was recently constructed in \cite{Grimm2} by blowing up a singular Fano threefold along a curve to produce a shrinkable brane worldvolume.  Following Donagi and Wijnholt \cite{Donagi} we can easily determine the Chern class of the normal bundle of \(S\) inside the ambient threefold \(X\) for all such models.  When the surface \(S\cong \mathbb{B}_{k}\mathbb{F}_{n}\) collapses, it does so by shrinking \(F\) and all of the \(E_{i}\) while keeping the curve \(B\) at finite size.  Grauert's criterion then tells us that \(c_{1}(N_{S/X})\) must be negative on the curves \(F\) and \(E_{i}\).  Further, for each \(i\) the cohomology class of \(F-E_{i}\) also represents a collapsed curve so \(c_{1}(N_{S/X})\) is also negative on this class.  Using the intersection ring \((\ref{intrela})\) it is easy to check that the solution to these constraints is:
\begin{equation}
c_{1}(N_{S/X})=-aB+(m-n-2)F+\sum_{i=1}^{k}c_{i}E_{i} \label{curven}
\end{equation}
Where the integers \(a\), \(m\), and  \(c_{i}\) in \((\ref{curven})\) are subject to the relation \(a>c_{i}>0\).  Meanwhile equation \((\ref{tensolve})\) together with the form \((\ref{cherns})\) of the Chern class of \(S\) tells us that class of the \(\mathbf{10}\) curve is:
\begin{equation}
\Sigma_{\mathbf{10}}=c_{1}(S)+c_{1}(N_{S/X})=(2-a)B+mF+\sum_{i=1}^{k}(1-c_{i})E_{i} \label{ntensolve}
\end{equation}
The class of \(\Sigma_{\mathbf{10}}\) must be a curve in \(S\), so in particular this means that in \((\ref{ntensolve})\) the coefficient of \(B\) must be non-negative, \(2\geq a\).  Combined with our previous inequality this implies \(a=2\) and \(c_{i}=1\) hence:
\begin{equation}
\Sigma_{\mathbf{10}}=mF \label{nntensolve}
\end{equation}
Notice that for a generic model where \(\Sigma_{\mathbf{10}}\) is an irreducible curve, \(m=1\) and \(\Sigma_{\mathbf{10}}\) is indeed a \(\mathbb{P}^{1}\) in agreement with our more general result \((\ref{ineq2})\).  
\paragraph{}
Thus we learn that \(SU(5)\) GUT models with a decoupling limit corresponding to an elementary contraction to a curve are characterized topologically by three natural numbers: \(n\) and \(k\) tell us that \(S \cong \mathbb{B}_{k}\mathbb{F}_{n}\) while \(m\) tells us the class of the \(\mathbf{10}\) curve via \((\ref{nntensolve})\).  In particular a choice of these three numbers uniquely fixes the number of Yukawa points via \((\ref{so12})-(\ref{su7})\):
\begin{eqnarray}
p(SO(12)) & = & 4m \\
p(E_{6}) & = & 2m \\
p(SU(7)) & = & 240+58m-30k 
\end{eqnarray}
\paragraph{}
Although the mathematical properties of these models can be described succinctly and have been discussed by several authors in the recent F-theory literature \cite{Grimm} \cite{Grimm2} \cite{Donagi}, it is unclear to us whether the decoupling condition of shrinking only to a curve and not to a point is really physically well-behaved.  To understand our skepticism the reader should recall from Section 3 that the physical decoupling limit of interest \emph{is not} the limit where \(S\) shrinks but rather it is the limit where the ambient threefold becomes very large while \(Vol(S)\) remains fixed.  Indeed only in the latter case does the Planck mass tend to infinity.  Although we have analyzed the properties of decoupling limits by working with a rescaled K\"{a}hler class we should take care that the geometry is under control when we rescale back to the physical metric.   To address this subtlety in detail let us denote by \(\omega_{phys}(t)\) the one-parameter family of K\"{a}hler classes relevant to the physical decoupling limit.  As \(t\rightarrow \infty \) we have:
\begin{equation}
\omega_{phys}^{3}(t)\sim Vol(X)(t)\sim M^{2}_{P}(t)\rightarrow \infty \hspace{.5in}\omega_{phys}^{2}(t)\cdot S \sim Vol(S)(t) \rightarrow \mathrm{finite} \neq 0 \label{physdec}
\end{equation}
Meanwhile, in contrast to the behavior \((\ref{physdec})\), we have the mathematically convenient K\"{a}hler class \(\omega_{math}(t)\) utilized throughout the later half of Section 3.  For the case of an elementary contraction to a curve we have for large \(t\):
\begin{equation}
\omega^{3}_{math}(t)\rightarrow \mathrm{finite} \neq 0 \hspace{.5in} \omega^{2}_{math}(t)\cdot S \rightarrow 0 \hspace{.5in}\omega_{math}(t)\cdot B \rightarrow \mathrm{finite} \neq 0 \label{mathdec}
\end{equation}
But now \(\omega_{phys}(t)\) differs from \(\omega_{math}(t)\) simply by an overall rescaling by a positive non-zero function of \(t\).  It then follows from \((\ref{physdec})-(\ref{mathdec})\) that parametrically for large \(t\):
\begin{equation}
\omega_{phys}(t)\cdot B\sim Vol(B)(t) \sim Vol(X)^{1/3}(t)\rightarrow \infty \label{blarge}
\end{equation}
This result is intuitively obvious.  If the size of \(B\) could be set independently from the size of \(X\), then \(S\) would admit a decoupling limit where it contracted to a point.  
\paragraph{}
Equation \((\ref{blarge})\) shows that decoupling limits which in the frame of \(\omega_{math}(t)\) are characterized by contractions to curves do not act trivially on \(S\) in the metric specified by \(\omega_{phys}(t)\).  Since the volume of \(S\) is fixed in the physical decoupling limit while the volume of \(B\) grows large, it must be that other curves in \(S\) become very small to compensate.  Thus asymptotically in the physical decoupling limit, \(S\) appears roughly as a very long and thin tube in \(X\).  It seems likely to us that due to these small curves in \(S\), large curvature corrections enter the seven-brane action and we lose control.  Equally strange, although these decoupling limits do achieve a parametrically small parameter \(M_{GUT}/M_{P}\), it does not appear that the zeroth order term in an expansion in this parameter is well defined.  In the strict decoupling limit, \(B\) has infinite volume, the corresponding small curves have zero volume, and at least as a four-dimensional quantum field theory with finite coupling constants, the seven-brane gauge theory does not make sense.  
\paragraph{}
It seems then that in the strongest sense of decoupling gravity while preserving the seven-brane gauge theory on \(S\), these models fail.  If one cannot consistently impose the physical decoupling limit then it does not make sense to use the mathematical criterion of shrinkablility to privilege these models over the most general F-theory GUT and in this regard these models are not on the same footing as those where \(S\) can shrink to a point.  Nevertheless, we have included them in the analysis throughout this paper because they illustrate an potentially interesting way to weaken the decoupling limit hypothesis.  Perhaps for certain examples the curvature corrections to the brane action in these models can be circumvented, in which case these decoupling limits seem to have a parametric separation of gauge and gravitational scales but admit no naive expansion in \(M_{GUT}/M_{P}\).  Understanding the precise physical implications of this scenario remains an open problem.
\subsubsection{A Non-Elementary Contraction to a Point}
\paragraph{}
The remaining viable possibility is then a seven-brane which in the frame of \(\omega_{math}(t)\) can undergo a non-elementary contraction to a point.  An example of this type was constructed in \cite{Mar}.\footnote{In an early version of \cite{Mar} the authors claimed that their threefold was Fano.  In agreement with our no-go result in Section 3, this is in fact not the case.  This in no way effects the rest of their work.}  The technique of their construction is similar in spirit to the proposed counterexample constructed in Section 3.  One begins with a Fano threefold which is the base of an elliptic Calabi-Yau fourfold and modifies the Fano by a sequence of blowups.  The characteristic feature of these constructions is that there is another surface \(S'\) which meets the \(SU(5)\) GUT brane \(S\) and also shrinks during the decoupling limit.  Intuitively, one strongly suspects that this \(S'\) will influence the local physics on \(S\).  There are at least two ways in which this might occur:
\begin{itemize}
\item A seven-brane might wrap \(S'\).  In this case we would have a gauge group which is a product \(G_{S'}\times SU(5)\), with the interesting feature that the coupling constant of the new gauge group scales parametrically with the GUT coupling:
\begin{equation}
 \alpha_{S'}\sim \frac{1}{Vol(S')}\sim \frac{1}{Vol(S)}\sim \alpha_{GUT}
\end{equation} 
\item A three-brane might wrap \(S'\).  In this case one expects an instanton contribution to the superpotential for the fields the GUT brane.  Although such contributions are exponentially suppressed, the model \cite{brokensusy} shows that such instantons can sometimes be the leading contribution to supersymmetry breaking.  
\end{itemize}
\paragraph{}
In fact we can make a more precise statement about the necessity of one of these two options which holds at least for the vast majority of cases.  To do this we will make use of Witten's characterization of three-brane instantons that contribute to the superpotential in F-theory  \cite{witten}.\footnote{Strictly speaking Witten's analysis applies only in the case where the three-brane meets no other branes, a condition which is explicitly violated here.  In the following we will be naive and assume that his results extend to this more exotic case.}   To deduce whether a three-brane wrapping \(S'\) contributes to the superpotential one considers not only \(S'\) but rather the threefold \(D\) obtained by restricting the Calabi-Yau fourfold to the part of the fibration over \(S'\).  Witten's result is that a sufficient condition for the three-brane to give a non-vanishing contribution to the superpotential is:
\begin{equation}
 \hspace{.5in} h^{1,0}(D)=h^{2,0}(D)=h^{3,0}(D)=0
\end{equation}
To analyze candidate three-brane instantons in our case we then need to constrain not only \(S'\), but also the characteristics of the elliptic fourfold near \(S'\).  
\paragraph{}
To proceed with the former first we consider the intersection \(S\cap S'\).  This is non-empty and contains a curve \(C\) on which the negativity of the normal bundle of \(S\) in \(X\) is violated, \(S\cdot C \geq0\).  Let us make the slightly stronger assumption that in fact \(S\cdot C \geq 1\).  Because the Mori cone of \(S\) is spanned by rational curves, we can then find a rational curve \(\Gamma\) also contained in \(S'\cap S\) with \(S \cdot \Gamma \geq 1\).  The normal bundle of \(\Gamma\) in \(X\) is split as:
\begin{equation}
N_{\Gamma/X}\cong N_{S/X}|_{\Gamma}\oplus N_{S'/X}|_{\Gamma}\cong N_{\Gamma/S'} \oplus N_{\Gamma/S}
\end{equation}
Hence the fact that \(S \cdot \Gamma \geq1\) means \(deg(N_{\Gamma/S'}\geq 1)\).  Thus \(S'\) is a surface which contains a rational curve which moves in a large family.  A result from the classification of algebraic surfaces then tells us that \(S'\) is itself a rational surface, related to \(\mathbb{F}_{n}\) by a sequence of blowups and blowdowns.  In particular \(h^{1,0}(S')=h^{2,0}(S')=0\).  Now to analyze the structure of the elliptic fibration restricted to \(S'\) we need only recall that \(S'\) is itself a shrinkable surface.  A straightforward application of the spectral sequence in \cite{Grassi} then implies that if there is no seven-brane wrapping \(S'\) then necessarily we have \(h^{i,0}(D)=0\) for \(i>0\) so a three-brane wrapping \(S'\) always contributes to the non-perturbative superpotential.\footnote{As with Witten's result Grassi's work \cite{Grassi} has assumptions which are violated in our example.  We will assume that the formalism developed there can be suitably generalized our more exotic setup.}
\paragraph{}
Thus we see that under very mild assumptions, non-elementary contractions are always accompanied by additional physics due to wrapped branes on the additional collapsing surfaces.  Understanding precisely the implications of this result, in particular how to compute the contribution of the three-brane instanton in this exotic situation then seems to be an important direction for future research. 
\begin{center}
\section{Conclusions}
\label{conclu}
\end{center}
\paragraph{}
One of the primary results of this paper is that there are a priori restrictions on the local singularities of compact elliptic Calabi-Yau fourfolds.  These appear in the form of compatibility conditions on the matter curves in a seven-brane and on the Yukawa couplings for these matter curves, and seem closely related to anomaly cancellation and the Green-Schwarz mechanism.  It is natural to suspect that the list of restrictions derived in this paper is not complete and that further effort might lead to new information.  Mathematically, the constraints on local singularities derived in this work all result from the observation that \(c_{1}(X)\) controls the locus of seven-branes in \(X\) and can be expressed at a brane in terms of only the local data of the singularity.    Phrased in this way, an obvious guess for a method to derive new constraints is then to study the the local behavior of the second Chern class \(c_{2}(X)\) near a seven-brane worldvolume.  For dimensional reasons it is natural to expect that \(c_{2}(X)\) has something to do with the matter curves in the compactification.  If a relation analogous to that of \(c_{1}(X)\) with the seven-branes exists then following the technique of Section 4 would likely give rise to interesting further restrictions on local configurations of seven-branes in any theory with gravity.
\paragraph{}
When combined with the assumption of a gravitational decoupling limit, the restrictions on local singularities derived in this paper yield powerful and phenomenologically interesting constraints on the form of local \(SU(5)\) GUTs.  Furthermore, the decoupling limit precludes the simplest class of UV completions, Fano threefolds.  In this regard another obvious and ambitious direction to pursue is to classify all threefolds \(X\) which can form the base of elliptic Calabi-Yau fourfolds.  The analogous problem for Calabi-Yau threefolds has been carried out, and the fact that birational geometry in three dimensions is well-understood suggests that such a classification might be tractable with existing technology.  A particularly relevant aspect of this classification for local F-theory GUTs is to understand how far from Fano \(X\) can be.  In Section 2 we have shown that the negativity of \(c_{1}(X)\) at a seven-brane is closely related with the seven-brane moduli.  In any complete model these moduli must all be stabilized and it would be interesting to understand how much of this can be achieved simply by requiring a decoupling limit.
\paragraph{}
Finally, in terms of the immediate physical applications of our work there is the obvious question of understanding what aspects of the phenomenologically attractive local models of Vafa, Heckman, et. al. can be dovetailed with the restrictions on local models derived in this paper.  The anomaly and Yukawa constraints \((\ref{constraint1}), (\ref{so12}), (\ref{e6}), (\ref{su7})\) imply that in order to consistently couple these models to gravity one must first solve the factorization problem discussed in Section 4.1.  Regardless of whether or not this issue can be overcome for their exact models, the results of this work suggest that there is reason to be hopeful about the prospects for constructions close in spirit to their ideas.  Indeed we have seen that a number of phenomenologically desirable ingredients discussed in \cite{M1} \cite{M2} \cite{M3} \cite{M4} \cite{brokensusy} such as enhanced Yukawa structures, three-brane instantons, and \(U(1)\) hidden sectors seem to be necessary properties of any consistent seven-brane GUT model with a decoupling limit.  What the results of this paper demonstrate is that these phenomenological components are tightly constrained and tied together in surprising ways.  The existence of these new restrictions together with the increasing proximity of upcoming collider experiments is likely to make the next stage of research on seven-brane GUTs a particularly exciting time.
\paragraph{}
\section*{Acknowledgments}  I would like to thank D. Simmons-Duffin, F. Denef, C. Vafa, J. Heckman, A. Tomasiello, and S. Katz for useful discussions and encouragement.  Additionally I would like to thank the Seventh Simons Workshop in Mathematics and Physics for hospitality during the latter stages of this project.
\appendix
\section{Limits of K\"{a}hler Degenerations}
In this appendix I will explain exactly how one can pass from a degeneration of a threefold \(X\) where a surface \(S\) shrinks to zero size in the sense of:
\begin{equation}
\lim_{t\rightarrow \infty} \hspace{.2in} \frac{(\int_{S}\omega_{t}^{2})^{3/4}}{(\int_{X}\omega^{3}_{t})^{1/2}}=0 \label{decoupled}
\end{equation}
to a compact limit geometry \(\hat{X}\) where \(S\) has degenerated to a curve or a point.  We will first consider the simplest case of a Fano threefold, and then later discuss generalizations to \(X\) which are non-Fano.
To proceed with the analysis, suppose \(S \subset X\) is such that it admits a decoupling limit as defined by \((\ref{decoupled})\).  Then by definition this means that there exists a of sequence of K\"{a}hler classes \(\omega_{n}\) for \(n= 1, 2, \cdots\) with the property that:
\begin{equation}
\lim_{n\rightarrow \infty} \ \  \frac{(\int_{S}\omega_{n}^{2})^{3/4}}{(\int_{X}\omega_{n}^{3})^{1/2}}= 0 \label{kahler}
\end{equation}  
Written in this form, it obvious that the decoupling condition is insensitive to the overall normalization of the K\"{a}hler form.  Given any sequence \(\{\omega_{n}\}\) satisfying \((\ref{kahler})\), we can obtain another such sequence by multiplying each \(\omega_{n}\) by any positive real number \(f_{n}\).  We will find it convenient to analyze the geometry of the decoupling limit by renormalizing the \(\omega_{n}\) such that a limiting class \(\omega\) exists.  To see that this is always possible we simply note that the cohomology \(H^{1,1}(X,\mathbb{R})\) where \(\omega_{n}\) takes values is a finite dimensional vector space.  We can fix a norm \(||\phantom{\omega}||\) on this vector space and replace \(\omega_{n}\) by \(\omega_{n}/||\omega_{n}||\).  It follows that our new sequence of K\"{a}hler classes has unit norm for all \(n\) hence as \(n\rightarrow \infty\) we can find a subsequence which converges to a nonzero class \(\omega\).  Notice that by the decoupling limit condition \((\ref{kahler})\) the limiting class \(\omega\) necessarily collapses the surface \(S\) to zero volume, so \(\omega\) lives in the boundary of the K\"{a}hler cone.  Depending on the resulting limit, the geometry of \(X\) might also degenerate.  For example, it could easily happen that \(\omega^{3}=0\) so that our original threefold looks asymptotically like a surface or curve.    
\paragraph{}
It is at this point that we can use the Fano condition to make the first of several simplifications.  We recall that on \(X\) one has the finite dimensional vector space spanned by numerical equivalence classes of curves.  Inside this vector space is a cone \(NE(X)\) which contains the effective curves.  The Mori cone theorem \cite{ConeThm} tells us the structure of the piece of this cone which intersects negatively with the canonical divisor of \(X\).  Since \(X\) is Fano every curve intersects negatively with \(K_{X}\), hence in this case the cone theorem yields complete information and \(NE(X)\) is of the form:
\begin{equation}
NE(X)=\sum_{i=1}^{m} \mathbb{R}_{+}[\Gamma_{i}] \label{cone}
\end{equation}
Where in the above \(\Gamma_{i}\) denotes a rational curve and the notation simply means that \(NE(X)\) is the convex hull of the rays generated by theses extremal curves.  In particular we see that the cone of curves is closed and hence its boundary \(\partial NE(X)\) can be defined by integral equations in an integral basis for \(H_{2}(B,\mathbb{Z})\).  Now, via the intersection paring on \(X\) we can view the K\"{a}hler cone, \(A(X)\), as the dual cone to \(NE(X)\).  Hence the boundary of this cone, where our limiting K\"{a}hler class \(\omega\) takes values, is also cut out by integral equations and therefore on the boundary of the K\"{a}hler cone the rational cohomology classes are dense.  That is \(H^{1,1}(X,\mathbb{Q}) \cap \partial A(X)\) is dense in \(\partial A(X)\).   We can therefore pick a sequence of rational classes \(\omega_{k} \in H^{1,1}(X,\mathbb{Q}) \cap \partial A(X)\) such that as \(k\rightarrow \infty \) the \(\omega_{k}\) approach our original limiting class \(\omega\).  Furthermore we can assume that for all \(k\), and any curve \(C\) on \(X\),  \(\omega_{k}\cdot C =0\) if and only if \(\omega\cdot C =0\), the key again being that these are integral equations on \(\partial A(X)\).  Since for our purposes, the only interesting information contained in \(\omega\) is exactly those curves on \(X\) which are collapsed, i.e. satisfy \(\omega \cdot C=0\) we see that we can assume that \(\omega\) is in fact a rational cohomology class.  Moreover, since the set of curves on \(X\) which are contracted is invariant under rescaling of the class \(\omega\) we can multiply \(\omega\) by a suitable positive integer and assume that it is an integral cohomology class (henceforth also called \(\omega\)).  We can therefore pick a line bundle \(\mathcal{L}\) with first Chern class \(\omega\) and study the K\"{a}hler degeneration of \(X\) via the geometry of \(\mathcal{L}\).
\paragraph{}
Now, as we have seen above the line bundle \(\mathcal{L}\) is non-trivial and since \(\omega\) lies in the boundary of the K\"{a}hler cone, \(\mathcal{L}\) has non-negative degree on every curve in \(X\). Thus in particular \(\mathcal{L}\) admits holomorphic sections.  Let  \(s_{1},\ldots, s_{n+1}\) denote a basis of these sections.  Then we can define a rational map from \(X\) to \(\mathbb{P}^{n}\) by:
\begin{equation}
b \mapsto [s_{1}(b):\cdots: s_{n+1}(b)] \label{linear system}
\end{equation}
If \(X\) were an arbitrary threefold then this map would not in general be holomorphic on all of \(X\) since it is ill-defined on the common vanishing locus of all of the sections \(s_{i}\).  However a theorem due to Kawamata \cite{BPFThm} tells us that on a Fano variety this complication does not occur, provided we pass to a sufficiently high power of the line bundle \(H\).  Applying this in our case we learn that some multiple \(m\) of our limiting class \(\omega\) determines a morphism from \(f: X \rightarrow \mathbb{P}^{n}\) for some \(n\) whose image will henceforth be denoted \(\widehat{X}\).  Since we are working up to scale on \(\omega\) we may as well assume that \(m\) is one.  By the decoupling condition \((\ref{decoupled})\) the image of our distinguished surface \(S\) has zero volume, and must be either a curve or a point.  Thus we see that the fact that \(X\) is Fano allows us to find a holomorphic map \(f: X \rightarrow \widehat{X}\) which carries out the given K\"{a}hler degeneration all at once, and that a necessary condition for decoupling gravity on \(S\) is that \(f(S)\) has dimension less than two.
\paragraph{}
Now let us attempt to generalize this lemma to the case where \(X\) is no longer assumed to be Fano.  As discussed in Section 3 the most interesting case is when \(\omega^{3}>0\) so that the limit of the threefold has non-zero volume, and from now we restrict to these examples.  In mathematical terminology this means that the map \(f\) defined above is a \emph{birational morphism}, i.e. a local modification of the threefold \(X\).  The main difficulty in generalizing our argument is that it is no longer true in general that the rational cohomology classes are dense on the boundary of the K\"{a}hler cone.  However if \(X\) is Calabi-Yau, a theorem due to Wilson \cite{cycone} implies that at least when \(\omega^{3}>0\) we can pass to a rational class as above.  Kawamata's theorem goes through and again we find a holomorphic map carrying out the limit of our K\"{a}hler degeneration.  More generally as long as \(c_{1}(X)\geq 0\) our argument goes through unmodified \cite{private}.  Finally, though we will not address this in the present work, we believe that a generalization of this idea should apply to all \(X\) which can form the base of an elliptically fibered Calabi-Yau fourfold.
\section{Negativity of Contraction}
Here we discuss the precise implications of the requirement of contractibility on a surface \(S\subset X\).  We have already argued intuitively in Section 3.1 that if \(S\) can shrink inside \(X\) then necessarily the normal bundle of \(S\) in \(X\) admits no holomorphic sections.  To understand this formally as well as to extract more detailed information, let us first study the case of a curve \(L\) which can shrink inside a surface \(D\) while leaving \(D\) at finite volume.  We expect that \(L\) should be rigid which means that its self-intersection number, \(L\cdot L\), is negative.  To see this we think of the degenerated limit of \(D\) inside an ambient projective space \(\mathbb{P}^{N}\) and consider a hyperplane section \(H\) of this limit.  Pulling back \(H\) to \(D\) itself we then obtain an ample divisor \(\widetilde{H}\) that satisfies the following self-evident intersections:
\begin{equation}
\widetilde{H}\cdot L =0 \hspace{.5in} \widetilde{H}\cdot \widetilde{H}>0
\end{equation}
Now we apply the Hodge index theorem.  The intersection form on the K\"{a}hler surface \(D\) has exactly one positive eigenvalue with an eigenspace spanned by \(\widetilde{H}\).  The orthogonal complement of \(\widetilde{H}\), which includes \(L\), therefore has negative-definite self-intersection numbers.  
\paragraph{}
To upgrade this argument to a shrinkable surface \(S\) inside a threefold \(X\) we simply take a generic hyperplane section \(H\) of the entire configuration.  In this way we find a curve \(C=S\cap H\) inside the surface \(H\) which can shrink while leaving \(H\) at finite two-dimensional volume.  Applying the previous analysis for shrinkable curves we learn that \(C\) has negative normal bundle in \(H\):
\begin{equation}
N_{C/H}\cdot C <0 \label{negc1}
\end{equation}
But since \(C\) is the transverse intersection of \(S\) and \(H\) its normal bundle is split as:
\begin{equation}
N_{C/X}\cong N_{C/H}\oplus N_{C/S}\cong N_{S/X}|_{C}\oplus N_{H/X}|_{C} \label{negc2}
\end{equation}
Hence combining \((\ref{negc1})-(\ref{negc2})\) we learn that along the curve \(C\), the normal bundle of \(S\) in \(X\) has negative degree, \(c_{1}(N_{S/X})\cdot C <0\).  Furthermore, by considering different hyperplane sections of \(S\) we see that the curve \(C\) can deform in \(S\) so \(deg(N_{C/S})\geq 0\).  This is the more precise statement we have been looking for.  If \(S\) can shrink then necessarily \(S\) contains a deformable curve \(C\) on which \(N_{S/X}\) has negative degree. This well known mathematical result is called \emph{negativity of contraction}.  It clearly implies that \(S\) is rigid.  Indeed if \(\sigma\) were a holomorphic section of \(mN_{S/X}\) with \(m>0\) then the locus in \(S\) where \(\sigma\) vanishes represents the Chern class of \(N_{S/X}\) and hence meets \(C\) negatively.  On the other hand, we can express the vanishing set of \(\sigma\) as:
\begin{equation}
\sigma=kC+E \label{ssplit}
\end{equation} 
Where in equation \((\ref{ssplit})\), \(E\) is a positive sum of curves distinct from \(C\) and \(k\geq0\).  Now intersect equation \((\ref{ssplit})\) with \(C\).  Since \(C\) and \(E\) are distinct complex manifolds, they meet non-negatively.  And since \(C\) can deform in \(S\) we have \(C\cdot C =deg(N_{C/S}) \geq0\).  Thus we deduce that \(\sigma \cdot C \geq0\) contradicting the fact that \(N_{S/X}\) has negative degree on \(C\).  We conclude that no positive power of the normal bundle of \(S\) in \(X\) admits holomorphic sections.
\section{Classification of Contractible Surfaces Inside Fano Threefolds}
\label{classification}
In this appendix we prove the classification result stated in section 3.  We will freely use the mathematical language and techniques of birational geometry.  For background material see \cite{ConeThm}.
\paragraph{}
\textbf{Theorem:}  Let \(f: X \rightarrow \widehat{X}\) be a birational morphism with \(X\) a smooth Fano threefold. Then \(f\) maps every smooth non-minimal surface \(S\subset X\) other than the Hirzebruch surface \(\mathbb{F}_{1}\) to a variety of dimension two in \(\widehat{X}\).     
\paragraph{}
\textbf{Proof:}
We are going to need a refined version of Mori's classification of extremal rays of birational type, which is true for smooth Fano threefolds \cite{fanoref}.  In the following table \(D\) denotes the exceptional divisor of a \(K_{X}\) negative extremal ray contraction generated by a rational curve \(\Gamma\), and \(C_{\Gamma}\) denotes the associated contraction morphism.  Mori's classification tells us that \(C_{\Gamma}\) is the inverse of blowing up \(C_{\Gamma}(X)\) at \(C_{\Gamma}(D)\).
\begin{center}
{ \small \begin{tabular}{|c | l |c |}
\hline 
Type of \(\Gamma\)  & \multicolumn{1}{c|}{\(C_{\Gamma}\) and \(D\)} & \(\Gamma\)\\
\hline
\(E_{1a}\) & \(C_{\Gamma}(D)\) a smooth curve, and \(C_{\Gamma}(X)\) is a smooth Fano threefold & \(\Gamma\) a \(\mathbb{P}^{1}\) fiber of \(D\) \\
& \(D\) a \(\mathbb{P}^{1}\) bundle.  & \(-K_{X}\cdot \Gamma =1\)\\
\hline
\(E_{1b}\) & \(C_{\Gamma}(D)\cong \mathbb{P}^{1}\), and \(C_{\Gamma}(X)\) a smooth threefold & \(\Gamma\) a \(\mathbb{P}^{1}\) fiber of \(D\)\\
&  \(D\cong \mathbb{P}^{1}\times\mathbb{P}^{1}\) with normal bundle  \(\mathcal{O}_{\mathbb{P}^{1}\times \mathbb{P}^{1}}(-1,-1)\) & \(-K_{X}\cdot \Gamma =1\) \\
\hline
\(E_{2}\) & \(C_{\Gamma}(D)\) is a point, \(C_{\Gamma}\)(X) is a smooth Fano threefold & \(\Gamma\) a line in \(D \cong \mathbb{P}^{2}\)\\
  & \(D \cong \mathbb{P}^{2}\) with normal bundle \(\mathcal{O}_{D}(D)\cong \mathcal{O}_{\mathbb{P}^{2}}(-1)\) & \(-K_{X}\cdot \Gamma =2\)\\
\hline
\(E_{3}\) & \(C_{\Gamma}(D)\) is an ordinary double point on \(C_{\Gamma}(D)\) & \(\Gamma\) either \(\mathbb{P}^{1}\) fiber of \(D\) \\
& \(D\cong \mathbb{P}^{1}\times \mathbb{P}^{1}\) with normal bundle \(\mathcal{O}_{D}(D)\cong \mathcal{O}_{\mathbb{P}^{1}\times \mathbb{P}^{1}}(-1,-1)\) & \(-K_{X}\cdot \Gamma =1\)\\
\hline
\(E_{4}\) & \(C_{\Gamma}(D)\) is a double point on \(C_{\Gamma}(X)\), \(D \cong\)quadric cone in \(\mathbb{P}^{3}\) & \(\Gamma\) a ruling of the cone \(D\)\\
& D has normal bundle \(\mathcal{O}_{D}(D)\cong \mathcal{O}_{D}\otimes \mathcal{O}_{\mathbb{P}^{3}}(-1)\)& \(-K_{X}\cdot \Gamma =1\) \\
\hline
\(E_{5}\) & \(C_{\Gamma}(D)\) is a quadruple point on \(C_{\Gamma}(X)\) & \(\Gamma\) a line in \(D \cong \mathbb{P}^{2}\) \\
& \(D \cong \mathbb{P}^{2}\) with normal bundle \(\mathcal{O}_{D}(D)\cong \mathcal{O}_{\mathbb{P}^{2}}(-2)\) &\(-K_{X}\cdot \Gamma =1\) \\
\hline
\end{tabular} }
\end{center}
\paragraph{}
Suppose that \(S\) is a smooth non-minimal surface other than \(\mathbb{F}_{1}\) which is contracted by the morphism \(f\).  We know from the above table that \(S\) is not contacted primitively.  Therefore \(S\) is contained in the linear span of the primitive exceptional divisors contracted by \(f\):
\begin{equation}
S=\sum_{i}a_{i}D_{i} \label{zsum}
\end{equation}
Our first lemma tells us the type of one of the rays:
\paragraph{}
\textbf{Lemma 1}:  At least one of the extremal curves \(\Gamma_{i}\) is of type \(E_{1a}\).
\paragraph{}
\textbf{Proof}:   Suppose that all rays are not of type \(E_{1a}\).  Observe that via the classification of extremal rays all of the exceptional divisors \(D_{i}\) has an ample conormal bundle, \(-D_{i}|_{D_{i}}>0\).  From this we deduce that every curve \(C\) on such an exceptional divisor \(D_{i}\) is movable.  To see this we simply apply the genus formula together with adjunction:
\begin{equation}
0 \leq g(C)=1+\frac{K_{D_{i}} \cdot C}{2}+ \frac{C\cdot C}{2} = 1+\frac{(K_{X}+D_{i})\cdot C}{2}+\frac{C\cdot C}{2}
\label{movable}
\end{equation}
Since both \(-K_{X}\) and \(-D_{i}\) are ample on \(D_{i}\) equation \((\ref{movable})\) implies that \(C \cdot C  \geq 0\).  It follows from this  \(D_{i}\) does not meet \(D_{j}\) for \(i\neq j\).  For if \(C \subset D_{i}\cap D_{j}\) is some effective curve contained in the transverse intersection of \(D_{i}\) with \(D_{j}\) then the normal bundle of \(C\) in \(X\) can be decomposed as:
\begin{equation}
N_{C/X}\cong N_{C/D_{i}}\oplus N_{C/D_{j}} \cong D_{j}|_{C}\oplus D_{i}|_{C}
\label{normal}
\end{equation}
But then:
\begin{equation}
0 \leq deg(N_{C/D_{i}})=E_{j} \cdot C
\label{cont1}
\end{equation}
Which contradicts the fact that \(E_{j}\) has ample conormal bundle.  More generally the intersection may not be transverse but the conclusion that exceptional divisors which are both not of type \(E_{1a}\) do not meet clearly remains valid.  Now we return to equation \((\ref{zsum})\).  Since \(S\) is not equal to any of the exceptional divisors \(E_{i}\), \(S\) meets some curve \(C \subset E_{j}\) non-negatively.  But then:
\begin{equation}
0\leq S\cdot C = \sum_{i}a_{i}E_{i}\cdot C=a_{j}E_{j}\cdot C \label{asum} 
\end{equation}
Since \(E_{j}\) has ample conormal bundle we learn that \(a_{j}\) is non-positive for all \(j\), which contradicts the fact that \(S\) is effective. \hspace{.5in}    Q.E.D.
\paragraph{}
Thus let \(\Gamma_{1}\) denote the ray of type \(E_{1a}\), Mori theory tells us that we can factor the morphism \(f\) through the contraction of ray generated by \(\Gamma_{1}\).  Thus we have the diagram:
\begin{equation}
\begin{xy} 
(0,0)*+{X} ="a"; (20,0)*+{Y_{1}}="b"; (10,-15)*+{\widehat{X}}="c";
{\ar^{C_{\Gamma_{1}}}"a" ; "b"}; {\ar_{f} "a" ; "c"}; {\ar ^{g}"b" ; "c"};
\end{xy} 
\label{factor}
\end{equation}
Where in the above \(Y_{1}\) is a smooth Fano threefold.  Now consider the image of \(S\) inside \(Y_{1}\); either \(C_{\Gamma_{1}}(S)\) is a primitive exceptional divisor of the morphism \(g\) or it is not.  If not, then we can apply Lemma 1 again to find another ray \(\Gamma_{2}\) of type \(E_{1a}\).  Iterating this procedure, we then see that after a sequence of \(E_{1a}\) type ray contractions, \(S\) must map to a primitive exceptional divisor \(D_{\gamma}\) associated to an extremal rational curve \(\gamma\) inside some smooth Fano threefold \(Y_{n}\).  Thus we can factor \(f\) as:
\begin{equation}
\begin{xy} 
(0,0)*+{X} ="a"; (20,0)*+{Y_{1}}="b"; (40,0)*+{Y_{2}}="c"; (60,0)*+{\cdots}="d"; (80,0)*+{Y_{n}}="e"; (100,0)*+{W}="f"; (50,-15)*+{\widehat{X}}="g";
{\ar^{C_{\Gamma_{1}}}"a" ; "b"}; {\ar^{C_{\Gamma_{2}}}"b" ; "c"}; {\ar^{C_{\Gamma_{3}}}"c" ; "d"}; {\ar^{C_{\Gamma_{n}}}"d" ; "e"}; {\ar^{C_{\gamma}}"e" ; "f"};{\ar_{f} "a" ; "g"}; {\ar^{h} "f" ; "g"}; 
\end{xy} 
\label{factor}
\end{equation}
Where \(S\) is contracted in the sequence of maps along the top of the diagram by first mapping to a primitive exceptional divisor \(D_{\gamma}\) and is then collapsed by the primitive contraction \(C_{\gamma}:Y_{n}\rightarrow W\).
\paragraph{}
Now we want to reconstruct \(S\) by blowing up.  Inside each smooth Fano threefold \(Y_{k}\) is a smooth curve \(Z_{k}\), and \(Y_{k-1}\) is obtained from \(Y_{k}\) by blowing up along \(Z_{k}\).  The following lemma is very useful for analyzing this situation:
\paragraph{}
\textbf{Lemma 2 (Mori Mukai):}
\begin{itemize}
\item
Let \(C \subset Y_{k}\) be any curve such that \(-K_{Y_{k}}\cdot C=1\).  Then either \(C\) is disjoint from \(Z_{k}\), or \(C=Z_{k}\)
\item
Let \(C \subset Y_{k}\) be any curve such that \(-K_{Y_{k}}\cdot C=2\).  Then either \(C\) is disjoint from \(Z_{k}\), or \(C=Z_{k}\), or \(C\) meets \(Z_{k}\) transversally at a single point.
\end{itemize}
\paragraph{} 
\textbf{Proof:}
Assume that \(-K_{Y_{k}}\cdot C \leq 2\), that \(C\neq Z_{k}\), and that \(C\) meets \(Z_{k}\).  Let \(D\) denote the exceptional divisor of the blowup \(C_{\Gamma_{k}}: Y_{k-1}\rightarrow Y_{k}\).  Then the canonical bundles of the two Fano threefolds \(Y_{k-1}\) and \(Y_{k}\) are related by:
\begin{equation}
-C^{\ast}_{\Gamma_{k}}(K_{Y_{k}})-D=-K_{Y_{k-1}} \label{canbundleform}
\end{equation}
Let \(\widehat{C}\) denote the strict transform of \(C\) under the blowup.  Then intersecting both sides of \((\ref{canbundleform})\) with \(\widehat{C}\) we find:
\begin{equation}
-K_{Y_{k}}\cdot C-D\cdot \widehat{C}=-K_{Y_{k-1}}\cdot{\widehat{C}}>0 \label{canbunform2}
\end{equation}
Where on the right-hand-side, the inequality comes from the fact that \(Y_{k-1}\) is Fano.  By hypothesis \(C \neq Z_{k}\) and \(C\) meets \(Z_{k}\) so \(C\cdot D >0\).  Then equation \((\ref{canbunform2})\) forces \(-K_{Y_{k}}\cdot C =2\) and \(D\cdot \widehat{C}=1\) which proves the lemma.
\paragraph{}
Now we can use Lemma 2 to constrain the type of the ray \(\gamma\)
\paragraph{}
\textbf{Lemma 3:} The extremal curve \(\gamma\) is of type \(E_{2}\).  
\paragraph{}
\textbf{Proof:}  Suppose \(\gamma\) is not of type \(E_{2}\).  Then the associated exceptional divisor \(D_{\gamma}\) is covered by the deformations of the extremal ray \(\gamma\) and \(-K_{Y_{n}}\cdot \gamma =1\).  Suppose that \(Z_{n}\) meets \(D_{\gamma}\).  By Lemma 2, we then learn that \(Z_{n}\subset D_{\gamma}\).  Blowing up \(Z_{n}\) we then see that the proper transform, \(\widehat{D}_{\gamma}\), of \(D_{\gamma}\) is again isomorphic to \(D_{\gamma}\).   Furthermore, it is clear that the proper transform \(\widehat{D}_{\gamma}\) is again covered by curves \(C\) with \(-K_{Y_{n-1}}\cdot C=1\) so we can apply the same argument to the \(\widehat{D}_{\gamma}\) inside \(Y_{n-1}\).  Proceeding inductively we learn that \(S\) is isomorphic to \(D_{\gamma}\) a primitive exceptional divisor. \hspace{.5in}  Q.E.D.
\paragraph{}
Thus the ray \(\gamma\) must be of type \(E_{2}\).  From Mori's classification we learn that \(D_{\gamma}\) is a smooth \(\mathbb{P}^{2}\) and \(-K_{Y_{n}}\cdot \gamma =2\).  Say \(Z_{n}\) meets \(D_{\gamma}\).  By Lemma 2 either \(Z_{n}\subset D_{\gamma}\) or \(Z_{n}\) meets \(D_{\gamma}\) transversally at a single point.  In the former case we blowup and the proper transform of \(D_{\gamma}\) is unmodified.  In the latter case, the proper transform \(\widehat{D}_{\gamma}\) is isomorphic to the Hirzebruch surface \(\mathbb{F}_{1}\).  Furthermore \(\widehat{D}_{\gamma}\) is covered by the curves consisting of the proper transforms of lines through the point where \(Z_{n}\) meets \(D_{\gamma}\) together with the exceptional curve.  Applying the blowup formula \((\ref{canbundleform})\) it is easy to see that these curves meet the anticanonical divisor of \(Y_{n-1}\) once, hence by the argument of Lemma 3, \(S\) is actually isomorphic to \(\widehat{D}_{\gamma}\) which is a Hirzebruch surface \(\mathbb{F}_{1}\).  This completes the proof of the theorem.   
\paragraph{}
In fact because of the work of Cutcosky, \cite{Cutcosky} this theorem generalizes to the case of arbitrary singularities of \(S\) and Gorenstein singularities of \(X\) provided we allow for the additional possibility of \(S\) the rank three singular quadric cone in \(\mathbb{P}^{3}\).  Gorenstein singularities are the natural class of singularities for \(X\) to consider in F-theory.  Indeed if the singularities are worse then Gorenstein then the canonical divisor of \(X\) can never be represented by a line bundle so the Weierstrass model construction of the Calabi-Yau presumably does not make sense.


\addcontentsline{toc}{section}{References}
\begin{thebibliography}{99}
\begin{spacing}{0}
\bibitem{curio} B. Andreas, and G. Curio, \emph{From Local to Global in F-Theory Model Building}, arXiv: hep-th: 0902.4143.
\bibitem{BHVI} C. Beasley, J.Heckman, and C. Vafa, \emph{GUTs and Exceptional Branes in F-Theory-I}, arXiv: hep-th: 0802.3391v1.
\bibitem{BHVII} C. Beasley, J.Heckman, and C. Vafa, \emph{GUTs and Exceptional Branes in F-Theory-II}, arXiv: hep-th: 0806.0102v2.
\bibitem{geosing} M. Bershadsky, K. Intrilligator, S. Kachru, D. Morrison, V. Sadov, and C. Vafa, \emph{Geometric Singularities and Enhanced Gauge Symmetries}, arXiv: hep-th: 9605200.
\bibitem{Grimm} R. Blumenhagen, T. Grimm, B. Jurke, and T. Weigand, \emph{F-theory uplifts and GUTs}, arXiv: hep-th: 0906.0013.
\bibitem{Grimm2} R. Blumenhagen, T. Grimm, B. Jurke, and T. Weigand,\emph{Global F-Theory GUTs}, arXiv: hep-th: 0908.1784.
\bibitem{M1} V. Bouchard, J. Heckman, J. Seo, and C. Vafa, \emph{F-Theory and Neutrinos: Kaluza-Klein Dilution of Flavor Hierarchy}, arXiv: hep-th: 0904.1419.
\bibitem{bourjaily} J. Bourjaily, \emph{Effective Field Theories for Local Models in F-Theory and M-Theory}, arXiv: hep-th: 0905.0142.
\bibitem{Cutcosky} S. Cutcosky, \emph{Elementary Contractions of Gorenstein Threefolds}, Math. Ann. \textbf{280} (3) (1988) 521-525.
\bibitem{Wijn} R. Donagi and M. Wijnholt, \emph{Model Building with F-theory}, arXiv: hep-th: 0802.2969.
\bibitem{Donagi} R. Donagi, and M. Wijnholt, \emph{Higgs Bundles and UV Completion in F-Theory}, arXiv: hep-th: 0904.1218v1.
\bibitem{Grassi} A. Grassi, \emph{On Minimal Models of Elliptic Threefolds}, Math. Ann. \textbf{290} (1991).
\bibitem{Grauert} H. Grauert, \emph{Ueber Modifikationen und exzeptionelle analytische Mengen}, Math. Ann., \textbf{146} (1962), 331-368.
\bibitem{GH} P. Griffiths and J. Harris, \emph{Principles of Algebraic Geometry}, John Wiley \(\&\) Sons, Inc., New York, 1978. 
\bibitem{brokensusy} J. Heckman, J. Marsano, N. Saulina, S. Shafer-Nameki, and C. Vafa, \emph{Instantons and SUSY Breaking in F-theory}, arXiv: hep-th: 0808.1286.  
\bibitem{M2} J. Heckman, A. Tavanfar, and C. Vafa, \emph{Cosmology of F-Theory GUTs}, arXiv: hep-th: 0812.3155.
\bibitem{M3} J. Heckman, A. Tavanfar, and C. Vafa, \emph{The Point of \(E_{8}\) in F-Theory GUTs}, arXiv: hep-th: 0906.0581.
\bibitem{M5} J. Heckman, and C. Vafa, \emph{F-Theory, GUTs, and the Weak Scale}, arXiv: hep-th: 0809.1098.
\bibitem{M4} J. Heckman, and C. Vafa, \emph{Flavor Hierarchy from F-Theory}, arXiv: hep-th: 0811.2417.
\bibitem{genfour} B. Holdom, W.S. Hou, T. Hurth. M.L. Mangano, S. Sultansoy, G. Unel, \emph{Four Statements about the Fourth Generation}, arXiv: hep-ph: 0904.4698.
\bibitem{katz} S. Katz, and C. Vafa, \emph{Matter From Geometry}, arXiv: hep-th: 9606086.
\bibitem{BPFThm} Y. Kawamata, \emph{The cone of curves of algebraic varieties}, Ann. of Math. \textbf{119} (1984), 603-633.
\bibitem{Mar2} J. Marsano, N. Saulina, and S. Sch\"{a}fer-Nameki, \emph{Gauge Mediation in F-Theory GUT Models}, arXiv: hep-th: 0808.1571.
\bibitem{Mar} J. Marsano, N. Saulina, and S. Sch\"{a}fer-Nameki, \emph{F-Theory Compactifications for Supersymmetric GUTs} arXiv: hep-th: 0904.3932v2.
\bibitem{Mar3} J. Marsano, N. Saulina, and S. Sch\"{a}fer-Nameki, \emph{Monodromies, Fluxes, and Compact Three-Generation F-theory GUTs} arXiv: hep-th: 0906.4672.
\bibitem{private} J. McKernan, Private Communication.
\bibitem{ConeThm} S. Mori, \emph{Threefolds whose canonical bundles are not numerically effective}, Ann. of Math. \textbf{116} (1982), 133-176.
\bibitem{fano1} S. Mori and S. Mukai, \emph{Classification of Fano 3-folds with \(B_{2}\geq2\)}, Manuscripta Math.  \textbf{36} (1981), 147-162.
\bibitem{fano2} S. Mori and S. Mukai, \emph{Erratum: Classification of Fano 3-folds with \(B_{2}\geq2\)}, Manuscripta Math.  \textbf{101} (2003), 407.
\bibitem{MV} D. Morrison and C. Vafa, \emph{Compactification of F-theory on Calabi-Yau Threefolds II}, arXiv: hep-th: 9603161v2.
\bibitem{fanoref} A.N Parshin and I. Shafarevich (Eds.), \emph{Algebraic Geometry V: Fano Varieties}, Springer, New York, N.Y., 1999. 
\bibitem{Sadov} V. Sadov, \emph{Generalized Green-Schwarz Mechanism in F-theory}, arXiv: hep-th: 9606008.
\bibitem{Sen} A. Sen, \emph{F-Theory and Orientifolds}, arXiv: hep-th: 9605150.
\bibitem{ftheory} C. Vafa, \emph{Evidence for F-theory}, arXiv: hep-th: 9602022v1.
\bibitem{cycone} P. M. H. Wilson, \emph{The K\"{a}hler cone on Calabi-Yau threefolds}, Invent. math. \textbf{107} (1992), 561-583
\bibitem{witten} E. Witten, \emph{Non-Perturbative Superpotentials in String-Theory}, arXiv: hep-th: 9604030.
\bibitem{Yau} S. T. Yau, \emph{On the Ricci Curvature of a Compact K\"{a}hler Manifold and the Complex Monge-Amp\`{e}re Equation I.}, Communications in Pure and Applied Mathematics, \textbf{31} (1978), 339-411. 
\end{spacing}
\end{thebibliography}
\end{document}